\documentclass{optica-article}

\journal{opticajournal} 

\articletype{Research Article}

\usepackage{gensymb}

\begin{document}

\title{Optimization Algorithm-Assisted Nanostructure Engineering for Performance Enhancement of Thin-Film Solar Cells}

\author{Abyaz Karim,\authormark{1} Sajid Ahmed Chowdhury,\authormark{1} A. F. M. Afnan Uzzaman Sheikh, \authormark{1} Asif Al Suny, \authormark{1} and Mustafa Habib Chowdhury \authormark{1,*}}

\address{\authormark{1}Department of Electrical and Electronic Engineering, Independent University, Bangladesh, Plot 16, Block B, Aftab Uddin Ahmed Rd, Dhaka 1229, Bangladesh\\}

\email{\authormark{*}mchowdhury@iub.edu.bd}



\begin{abstract*} 
A computation-based study is presented that utilizes the finite-difference time-domain (FDTD) method to perform optical simulations. These simulations were used to observe the effects of core-shell nanostructures with triangular cross-sections embedded inside the light absorbing layer of thin-film solar cells (TFSCs), for improving the optoelectronic performance. Despite extensive research, the relationship between the geometric parameters of light trapping nanostructures and TFSC performance is not yet fully elucidated. To assist in finding the optimal set of parameters, optimization algorithms are used to find a set of parameters that provide maximal short-circuit current density (J\textsubscript{SC}). Three separate algorithms were applied, and the results were scrutinized in-depth with a sensitivity analysis, observing the effects of small changes in the optimal parameters. The optimal parameters of the nanostructure produced J\textsubscript{SC} values ranging from 28.90 to 35.39 mA/cm\textsuperscript{2} compared to 13.69 mA/cm\textsuperscript{2} for a reference TFSC with no nanostructure present. From the results of this study, it can be inferred that optimized periodic grating structures can provide substantial improvement in performance of TFSCs.
\end{abstract*}

\section{Introduction}
To meet worldwide energy demands amid depleting non-renewable resources and a rapidly growing global population, continuous research and development of solar photovoltaic (PV) technology is paramount \cite{kannan_solar_2016}. In fact, for the needs of future sustainable development, PV technologies and devices can be a reliable source of renewable energy \cite{maka_solar_2022}. Among the various types of PV technology and solar cells that have been developed, thin-film solar cells (TFSCs) have been a research focus due to their lower resource use and higher power-to-weight ratios. The appeal of TFSCs comes from the devices being created by deposition of material in thin films whose thicknesses are measured in the nano- and micrometer ranges \cite{elkhamisy_comprehensive_2024}. With such low thickness ranges, there is the obvious advantage of flexibility, reduced weight, and reduced material usage for manufacturing. Increasing power-to-weight ratio \cite{li_flexible_2024,panagiotopoulos_critical_2023} has been a targeted goal of a few studies because the application of efficient TFSCs extends beyond mass power generation, also applying to electronic devices such as solar power chargers, portable devices (smart watches, calculators, cell phones) and also for space applications such as satellites, data centers and drones \cite{otte_flexible_2006}.  

However, reduced thickness of the light absorbing layer is a major cause of poor efficiency in TFSCs \cite{poortmans_thin_2006}. To ameliorate this, light management or light trapping structures are used to increase absorption through various means, such as lowering reflectivity, which in turn produces more electron-hole pairs for photocurrent generation \cite{brongersma_light_2014}. Subwavelength-sized nanostructures or nanotextures can be engineered for light trapping within TFSCs. These techniques can be broadly categorized into: textured top-layers/anti-reflective coatings (ARCs), textured back-contacts/nano-gratings or scattering and energy transfer facilitated by plasmonic metal nanostructures \cite{haug_light_2015}. While simply adding engineered nanostructures can provide a higher baseline efficiency, studies have found that the final output of TFSCs depends significantly on the geometric parameters of these nanostructures \cite{nayeem_arefin_influence_2020,huang_thin-film_2022}. Thus, computational analysis of nanostructures of various morphologies can provide valuable insight to the manufacture of cost-effective and efficient TFSCs. However, with more complex structures which have a greater number of independent parameters (for example, a cuboid having three independent dimensions compared to a cube), the complexity of analysis scales up much faster.  Moreover, combinatorially analyzing parameter values within a specific range can be highly inefficient. As a result, a number of studies have been dedicated to searching for optimal parameters using metaheuristic techniques, otherwise known as optimization algorithms \cite{tsai_handbook_2023, Li2019}. Optimization algorithms or metaheuristic techniques have the objective of maximizing or minimizing an output parameter of a system, based on input parameters. As a result, favorable parameters that fulfill a given objective can be found within a large pool of possible parameter combinations. Moreover, parameter combinations that do not contribute to the desired minimization or maximization of an output can be overlooked with algorithm assistance, thus, saving computational resources.

Previous studies have demonstrated that optimization algorithms can be used to search for optimal or near-optimal input parameter values for light trapping structures in TFSCs \cite{karim_optimizing_2024,zhou_study_2013,haque_effects_2023,muhammad_optimization_2018}. This study intends to provide novelty by further analyzing the optimal solutions obtained. Optimal results can be tested for robustness by slightly changing the input parameters and observing the change in the desired output. Methods demonstrated in this study show that the sensitivity of the optimized solutions can be tested with finite-difference time-domain (FDTD) simulations. Sensitivity refers to how sharply the output of a system may change in response to perturbation in the input parameter(s). Zhou et al. \cite{zhou_study_2013} shows an example of this, where deviation in optimal nanoparticle shape resulted only in minor decrease in the performance of the TFSC.

\begin{figure}[b]
    \centering
    \includegraphics[width=12cm]{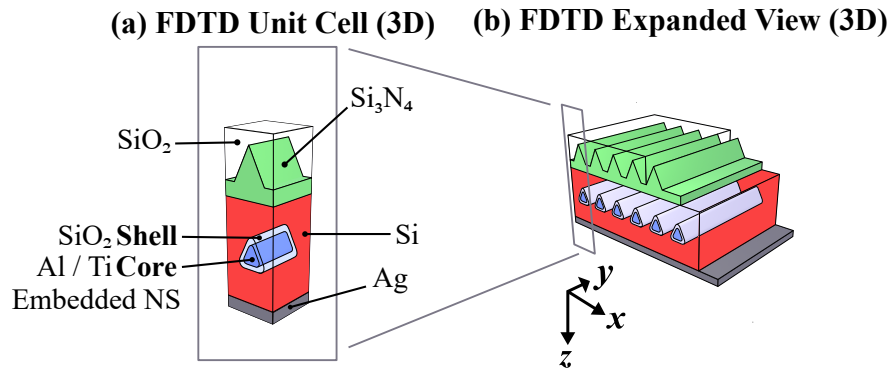}
    \caption{Model showing the design of the TFSC in 3D. (a) The simulated unit cell in FDTD. (b) An expanded view to reinforce that the design has a constant cross-section in the y-direction. The triangular core-shell nanostructure serves the function of a core-shell nanoparticle and a grating structure. Such a structure can contribute to improved TFSC performance through phenomena such as LSPR, light trapping and scattering.}
    \label{fig:3dExpanded}
\end{figure}

This study also investigates nanostructures modeled with fabrication-induced imperfections and geometric truncations to more accurately emulate realistic nanofabrication process variations. As a result, the analysis in this study can be robust from the perspective of TFSC design and optimization. The nanostructure investigated in this study can be likened to a nano-grating, except that it is not attached to the bottom layer of the TFSC. It consists of a metal core and dielectric shell; for example an aluminum (Al) core and silicon dioxide (SiO\textsubscript{2}) shell. Layer-wise structure of the TFSC and the embedded nanostructure have been shown in Fig. \ref{fig:3dExpanded} - illustrating how simulation of a unit cell can translate to a repeating structure in a TFSC. The reason for using a core-shell structure was to reduce the chances of recombination at the metal-semiconductor interfaces, to exploit phenomena such as localized surface plasmon resonance (LSPR) \cite{hasan_use_2021, Maier2007-xg} and induce the propagation of surface plasmon polaritons (SPPs) \cite{sultan_particle_2024}.

The input parameters used to control nanostructure geometry have a significant effect on the output of the TFSC. An optimization algorithm changes inputs through iterations to maximize or minimize a “figure of merit” (FOM) to fulfill its objective function (a mathematical definition of the goal of an optimization process). For this study, the FOM is the short-circuit current density (J\textsubscript{SC}) and the objective function is to maximize this value based on input parameters within given constraints \cite{sultan_particle_2024, karim_optimizing_2024}. To obtain a greater number of optimal solutions, results from the three algorithms: Particle Swarm Optimization (PSO), Differential Evolution (DE), and Dual Annealing (DA) were analyzed. Use of the above mentioned algorithms for TFSC and photonics related studies have already been documented in existing literature \cite{elsheikh_review_2019,kaya_extremely_2017,hara_optimum_1992,zhou_study_2013}. The physical input parameters of the nanostructure and the J\textsubscript{SC} output are not related explicitly through any formulae. Thus, multiple algorithms with different search strategies can increase the number of solutions obtained. This can be beneficial for finding more optimal configurations which can provide more choices for manufacturing, and allow further insight on the effects of each parameter. Overall, this study aims to provide a thorough analysis of TFSC design and performance enhancement with nanostructures using optimization algorithms.

\section{Modeling and Methodology}

\subsection{Simulation Setup}

\begin{figure}[t]
    \centering
    \includegraphics[width=12cm]{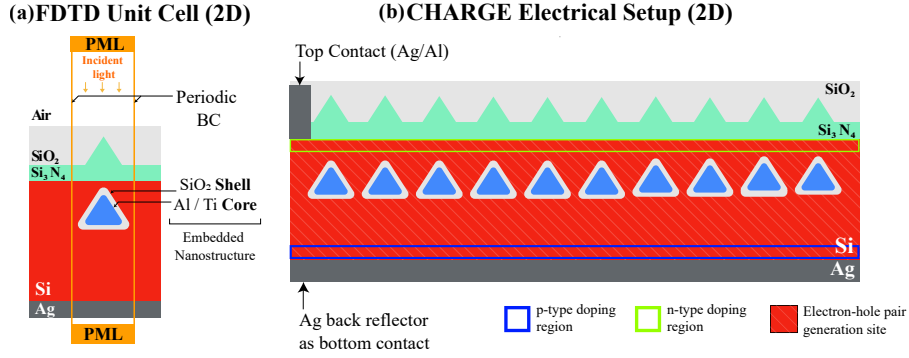}
    \caption{A 2D diagram representing the TFSC design being simulated in (a) FDTD and (b) CHARGE. A single unit cell is simulated in FDTD and is periodically repeated a given number of times to produce a solar generation rate matrix, which is transferred over to CHARGE to find a value of J\textsubscript{SC} affected by factors such as recombination and doping.}
    \label{fig:simSetup}
\end{figure}

All simulations were done using the commercially available Ansys Lumerical Software Suite. The main software used were finite-difference time-domain (FDTD) for optical simulations and CHARGE (3D charge transport solver) for the electrical simulations \cite{ansys_lumerical_suite}. Effects of heating through the movement of charge carriers like electrons, can be simulated with HEAT (Heat transport simulation). This study is solely based on computation and simulation, which can be broken up into two main parts. Firstly, the optical simulation is facilitated by FDTD producing electron-hole pair generation rate data and J\textsubscript{SC}. This data can then be transferred to CHARGE to produce the electrical simulation results. The CHARGE simulation results include: the final recalculated J\textsubscript{SC} value (considering factors such as doping and recombination), open-circuit voltage (V\textsubscript{OC}), fill factor (FF), maximum output power (P\textsubscript{max}) and power conversion efficiency $(\eta)$. Fig. \ref{fig:simSetup} shows the setup described for both types of simulation. When considering the temperature increase, initial heat production results from FDTD can be imported and with the HEAT simulation, temperature dependent electrical results can be produced \cite{planar_si_sc}. As shorthand, these temperature dependent results will be referred to as “thermo-electrical results”. 

The environmental conditions have been fixed at 1000 W/m\textsuperscript{2} solar irradiance at an air mass ratio of AM1.5G and an ambient temperature is 300 K \cite{nayeem_arefin_influence_2020}. The range of wavelengths of the incident light source ranges from 400 nm to 1100 nm. These conditions are applied to Fig. \ref{fig:simSetup}(a). The unit cell illustrated in Fig. \ref{fig:simSetup}(a) is the simulation region defined by the FDTD boundary conditions. There is a perfectly matched layer (PML) on the top and bottom that is designed to absorb all light, with periodic boundary conditions for the sides of the cuboidal region.  Furthermore, J\textsubscript{SC} is calculated from the electron-hole pair generation rate that is dependent on the quantum efficiency (QE), which in turn depends on the absorber substrate bandgap, which is 1.12 eV for silicon. The bottom layer of the TFSC is the 100 nm silver (Ag) back reflector which lies below a 1000 nm thick silicon absorbing substrate. The nanostructures that are being extensively studied are embedded within this silicon layer. On top of the silicon layer, the ARC consists of two layers, made of silicon nitride (Si\textsubscript{3}N\textsubscript{4}) and SiO\textsubscript{2}, respectively. This layer-wise structure is displayed in Fig. \ref{fig:3dExpanded} and Fig. \ref{fig:simSetup} from both 3D and 2D perspectives respectively. The Si\textsubscript{3}N\textsubscript{4} layer has a triangular cross-sectional conformation, embedded within a layer of SiO\textsubscript{2}. The height of the Si\textsubscript{3}N\textsubscript{4} triangle texture is dependent on the height of the triangular cross section of the embedded nanostructure, similar to a design involving pyramid texture in a study by Huang et al. \cite{huang_thin-film_2022}. This has been done to  prevent creating an additional independent input parameter to avoid further complexity. Keeping the height of the texture constant could interfere with other parameters that will be discussed. The materials making up the embedded nanostructures consist of a SiO\textsubscript{2} shell coating and two possible metal cores: aluminum (Al) or titanium (Ti). While Al is well-known for its plasmonic properties and applications in photonics \cite{shukla_aluminum_2025}, Ti is less covered in the literature. Yet, in studies where it has been featured, it is beneficial for its LSPR capabilities in the forms of nanofilms or back-gratings \cite{li_research_surfacePlasmon_2025,sultan2023TENCON}. 

The optimization of the geometric parameters of the nanostructure takes place during the optical simulations (FDTD) phase. PSO, DE or DA are implemented at this stage such that, according to an objective function, the parameters can be adjusted iteratively till a tolerance is achieved or a predetermined number of simulations are finished. Firstly, the optimization process with PSO is done through Ansys Lumerical FDTD’s built-in optimization utility \cite{optimization_utility}. A Python script is run alongside the FDTD simulation for the DE and DA optimizations. The source code for DE \cite{storn_differential_1997,qiang_unified_2014} and DA \cite{tsallis_generalized_1996} are drawn from the open-source Python library SciPy \cite{virtanen}. At the end of the optimization process, the set of input parameters produce a corresponding optimal output, which for this instance is the J\textsubscript{SC}. The electron-hole pair generation data can be extracted from this optimized solution to proceed with the electrical simulations. For additional optical analysis, FDTD simulations can also provide the absorption spectra and electric fields. The effects of heat on the output can also be simulated with a temperature profile from data from optical simulations \cite{planar_si_sc}. 

As seen in the setup in Fig. \ref{fig:simSetup}(b), the electrical simulations allow for a more accurate value for the J\textsubscript{SC} to be calculated by considering factors such as p-i-n doping configuration, electron/hole mobility and different types of electron-hole pair recombination. There are different types of recombination, including trap-assisted/defect assisted recombination, radiative recombination, and Auger recombination. Usually, the most dominating form is radiative recombination \cite{cb_honsberg_pveducation_nodate}. Additionally, trap-assisted recombination features an additional defect energy level (also known as a trap-state) between the conduction and valence band \cite{ma_trapassisted_2023}. Auger recombination involves recombination of a conduction band electron and valence band hole where the energy is released instead of producing a photon, which is transferred to another conduction band electron \cite{cb_honsberg_pveducation_nodate, Luo2016}. The electric properties of the solar cells are extracted via sweeping through a range of voltages across the top and bottom metal contacts. This allows finding the variation of the current density (J) and output power with voltage. The aggregate of the optical, electrical, and thermo-electrical results allows for a robust assessment of TFSC performance.

\subsection{Parameters and Constraints for Embedded Nanostructure}

\begin{figure}[b]
    \centering
    \includegraphics[width=12cm]{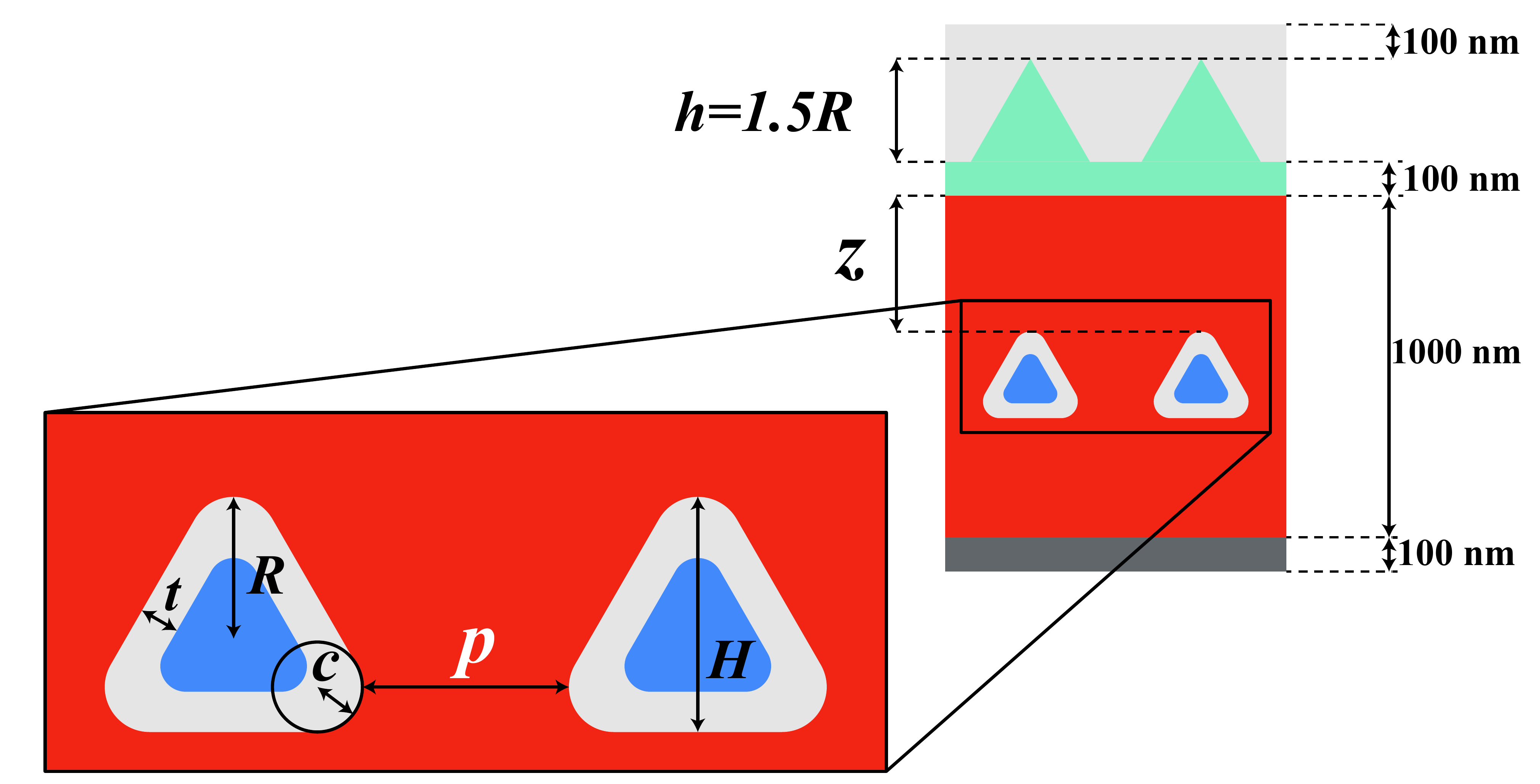}
    \caption{Relevant parameters to optimize shown on two adjacent unit cells. The parameter labeled h is made equal to $1.5R$ (equal to the height of the triangular cross-section of the embedded nanostructure assuming $c$ is 0). $H$ is the height of the rounded triangle cross-section. On the diagram the labeled parameters are as follows: $R$ - radius, $c$ - corner radius, $t$ - shell thickness, $p$ - periodicity and $z$ - vertical position.}
    \label{fig:parameters}
\end{figure}

The nanostructure in every unit cell has a rounded equilateral triangular cross section and is of a core-shell configuration. As mentioned previously, this periodically repeating structure has a function similar to a textured nano-grating \cite{sultan_particle_2024}. A dielectric shell of the nanostructure provides a barrier between the semiconductor and the metal core which is preferred for a reactive metal such as Al; even metals that are less reactive should be coated with a dielectric shell to prevent cases of recombination after electron-hole pairs are formed \cite{Hasan2021}. Moreover, a SiO\textsubscript{2} dielectric shell can be beneficial in red shifting the absorbance/scattering/extinction spectra, allowing better absorption of low energy photons in the near-infrared range and provide additional chemical stability \cite{nayeem_arefin_influence_2020}. Spherical metal-dielectric core-shell nanoparticles have been investigated and can provide enhancement through phenomena such as near-field enhancement, far-field scattering, hot electron transfer (HET) and plasmon resonant energy transfer (PRET) \cite{yu_effects_2017}. There are five parameters that control the size, geometry and position of the triangular nanostructures. Fig. \ref{fig:parameters} shows that the parameters are: $R$ – the radius which is the distance from the center to the corner of the cross section; $t$ – the SiO\textsubscript{2} shell thickness where it should be noted the radius of the core is $(R-t)$; $c$ – the corner radius is the roundness of the corners/edges of the nanostructure; $p$ – defines periodicity and is the shortest normal distance between two adjacent structures in an array; $z$ – the final parameter to control for the distance from the top of the triangle to the top of the silicon absorber layer surface (in the z-axis direction). Furthermore, in Fig. \ref{fig:parameters}, a value $h$ (height of the ARC triangle texture) can be seen labeled, which is made dependent on the value of $R$  and is not an independent parameter. If $h$ were to be kept constant, unintended designs could result. The height of the rounded triangle cross-section represented by $H$ and is equal to $0.5(3R+c)$.

\begin{figure}[t]
    \centering
    \includegraphics[width=10cm]{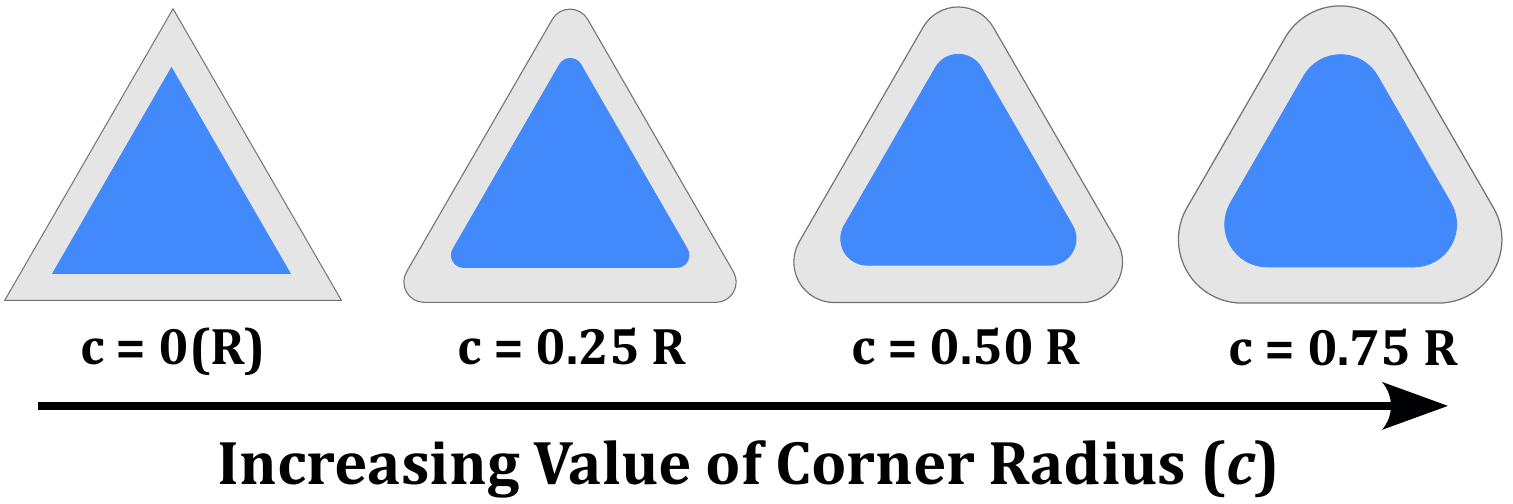}
    \caption{Illustration of the increasing value of the corner radius, $c$. The ceiling is set at 75\% of the value of $R$, to prevent unexpected design errors within simulations.}
    \label{fig:increasingCornerRad}
\end{figure}

Some parameters need to be subject to certain constraints to ensure that all possible designs of the TFSC are realistic and within expectations. The two parameters that will be subject to conditional constraints are the vertical position $(z)$ and the corner radius $(c)$.  The minimum value of $z$ is 0, which occurs only when the apex of the triangular nanostructure touches the top of the silicon layer. At the maximum value of z (lowest possible position), the base of the nanostructure touches the interface between the back reflector and the silicon absorber layer. Further details on the conditional constraints on the parameters and the bounds for $z$ can be found in the supplemental document illustrated with Fig. S1 and Fig. S2. As for the parameter c, a conditional ceiling is applied such that unwanted design errors are not produced. This condition is that the value of $c$ cannot be greater than 75\% of $R$ - this is illustrated in Fig. \ref{fig:increasingCornerRad}. The constraints for all the parameters are listed in Table \ref{tab:parameter_constraints}. 

\begin{table}[h]
    \centering
    \begin{tabular}{|l|l|}
        \hline
        \textbf{Parameters} & \textbf{Range} \\ \hline
        $R$ (Radius) & 50 nm $\leq R \leq 250$ nm \\ \hline
        $t$ (Shell Thickness) & 10 nm $\leq t \leq 70$ nm \\ \hline
        $c$ (Corner Radius)* & 0 nm $\leq c \leq 0.75R$ \\ \hline
        $p$ (Periodicity) & 0 nm $\leq p \leq 300$ nm \\ \hline
        $z$ (Vertical Position)* & 0 nm $\leq z \leq [1000-0.5(3R+c)]$ nm \\ \hline 
    \end{tabular}
    \caption{The upper and lower limits of the optimization range for each of the parameters; *c and $z$ are annotated here to indicate that they are subject to conditional constraints }
    \label{tab:parameter_constraints}
\end{table}

\section{Numerical Analysis}

\subsection{Equations and Calculations of Essential Figures}

Analysis in this study begins with the FDTD simulations that produce optical results based on a few equations. Here the relevant values such as the electric field and magnetic field are brought about through the solving of time-domain Maxwell equations within each spatial Yee cell \cite{Teixeira2023}. Firstly, power absorbed per unit volume ($P_{abs}$) is an essential value to calculate; specifically the power that is absorbed into the silicon absorber layer. $P_{abs}$ is calculated through the divergence on the Poynting vector \cite{solar_cell_methodology,moree_comparison_2022}. However, this version of $P_{abs}$ is prone to being unstable. A more stable formula is described in Eq. \ref{eq:Pabs_raw} \cite{su_based_2021}. Eq. \ref{eq:Pabs_raw} involves the frequency, magnitude-squared of the electric field, and imaginary component of permittivity:

\begin{equation}
P_{abs}=-0.5 \omega|E(r,\omega)|^2 \, \Im[\varepsilon(r, \omega)] 
\label{eq:Pabs_raw}
\end{equation}

$E$ and $\varepsilon$ are electric field and complex permittivity functions, which is why the imaginary component of the permittivity is specified here with $\Im[\varepsilon(r, \omega)]$. Furthermore, $\omega$ represents the angular frequency here. $P_{abs}$ is a part of the generation rate function per frequency, $G(r)$ which is defined as follows \cite{su_based_2021} in Eq. \ref{eq:genRate} below:

\begin{equation}
G(r) = \frac{P_{abs}}{\hbar \omega} = \frac{-0.5 \omega|E(r,\omega)|^2 \, \Im[\varepsilon(r, \omega)]}{\hbar \omega} = \frac{-0.5 |E(r,\omega)|^2 \, \Im[\varepsilon(r, \omega)]}{\hbar } 
\label{eq:genRate}
\end{equation}

It is apparent from observing the ratio described in Eq. \ref{eq:genRate}, that the denominator here represents the input photon energy; this leads to a cancellation leading to the final ratio, leaving behind the reduced Planck constant ($\hbar$). For a particular frequency, this gives the number of photons produced on the condition that the photon energy is above the bandgap energy, which in this study is 1.12 eV for silicon. The formula for $G(r)$ provides a matrix produced by the analysis of an FDTD simulation and transferred to the CHARGE simulation to generate electrical parameters after considering factors such as doping and recombination. Additionally, quantum efficiency $(QE)$ is an important metric for solar cell performance and is defined as the following ratio with Eq. \ref{eq:QE} \cite{solar_cell_methodology}, which can then be used to estimate J\textsubscript{SC} from FDTD simulations with Eq. \ref{eq:SCCD_Calc}:

\begin{equation}
QE(\lambda) = \frac{P_{abs}(\lambda)}{P_{in}(\lambda)}
\label{eq:QE}
\end{equation}

\begin{equation}
J_{SC} = e \int_{\lambda_1}^{\lambda_2} \frac{\lambda}{hc} \, QE(\lambda) I_{AM1.5G}(\lambda) \, d\lambda
\label{eq:SCCD_Calc}
\end{equation}

The constants for Eq. \ref{eq:SCCD_Calc} are $e$ – the elementary charge, $h$ and $c$ are Planck’s constant and speed of light respectively. The calculation of J\textsubscript{SC} involves the integration of the products of $QE$ and the intensity of incident radiation at AM1.5G ($I_{AM1.5G}$) both of which are functions of wavelength. The integration bounds are the wavelength range ($[\lambda_1,\lambda_2]$) which is 400-1100 nm. J\textsubscript{SC} calculated with Eq. \ref{eq:SCCD_Calc} is an effective initial measure of TFSC performance, given its dependency on $QE(\lambda)$, which is calculated in a similar way to generation rate \cite{solar_cell_methodology}. As the optimization process uses the optical simulations, the value calculated using Eq. \ref{eq:SCCD_Calc} can be an effective FOM. In addition, many of the other electrical parameters are directly dependent on J\textsubscript{SC}. However, the dependency on $QE(\lambda)$ will also cause this estimate of J\textsubscript{SC} to be higher. This is because it is assumed that every photon absorbed would lead to carrier generation. By accounting for all possible effects of recombination, the CHARGE electrical simulations can assess the effect on generation rate and calculate a more accurate value for J\textsubscript{SC} and other electrical performance metrics as well \cite{solar_cell_methodology, sultan_particle_2024, Suny2026}. 

Given that this series of calculations have been established for the FDTD simulation, a necessary objective function for the optimization algorithms can be formulated. As previously mentioned, nanostructure morphology, which is dependent on physical input parameters, plays a significant role in TFSC output. Thus, J\textsubscript{SC} can be described as a mathematical function dependent on the input parameters represented as a parameter vector $[R,t,c,p,z]$ as shown in Eq. \ref{eq:paramFunction}.

\begin{equation}
\begin{split}
parameters &= [R,t,c,p,z] \\
J_{SC} = F[parameters] &= F(R,t,c,p,z)
\end{split}
\label{eq:paramFunction}
\end{equation}

Furthermore, the objective function for the algorithms to fulfill can be mathematically described as in Eq. \ref{eq:ObjFunc}:

\begin{equation}
\begin{split}
\max_{R,t,c,p,z} \ J_{SC} &= \max_{R,t,c,p,z} \ F(R,t,c,p,z) \\
subject \ to & \quad [R,t,c,p,z]\ \in \ K 
\end{split}
\label{eq:ObjFunc}
\end{equation}

Here, the parameters are subject to being elements of set $K$. $K$ is defined by the constraints listed in Table \ref{tab:parameter_constraints}, and the parameter values that qualify for this set form the solution space that the algorithms (PSO, DE, and DA) will search. After the optical FDTD simulations, the absorption spectra are plotted and subsequently the relevant data is extracted. Absorption is determined by measuring the incident power entering and leaving the system within the simulation using power/field monitors. Generally, the following equation as a function of wavelength as shown in Eq. \ref{eq:ART} is used to plot an absorption spectra to analyze what wavelengths are being absorbed into the TFSC \cite{wang_design_2017}.

\begin{equation}
A(\lambda)= 1-R(\lambda)-T(\lambda)
\label{eq:ART}
\end{equation}

Once an algorithm has converged on a solution with favorable performance, the generation rate data and heat production data can be transferred to CHARGE. With CHARGE, the calculation of the remaining values needed to assess the electrical performance parameters for the TFSC is done. Firstly, the recalculated J\textsubscript{SC} accounting for the aforementioned factors is computed. Then, the open-circuit voltage (V\textsubscript{OC}), which is the maximum possible TFSC voltage is calculated. V\textsubscript{OC} can be measured across the contacts at zero current. Fill factor (FF) is a measure of how close the IV (current-voltage) curve of the solar cell is to ideal. The maximum power per unit area is labeled as P\textsubscript{max}. Lastly, power conversion efficiency ($\eta$), which is the ratio of output power to input power, is multiplied by 100\%. Formulae for calculating these parameters can be found in \cite{cb_honsberg_pveducation_nodate, su_based_2021, Suny2026}. The calculation of electrical performance parameters and the use of three optimization algorithms reflect the performance of the TFSC realistically. In addition to this, the use of multiple algorithms helps in searching for as many favorable parameter combinations as possible from the pool of all possible designs. Ideally, all algorithms should converge towards the same global maximum point (input parameters that yield maximum J\textsubscript{SC} values), but may however, differ because of different search methods. Moreover, in this study there are five inputs ($[R,t,c,p,z]$) and one output (J\textsubscript{SC}), which makes graphical visualization difficult. As a result, there can be chances that the algorithms may converge to solutions with similar (marginally different) J\textsubscript{SC} values but with varying input configurations (local maxima). Thus, by using more than one algorithm, these solutions can be compared. A more in-depth comparison of these solutions can be done by analyzing the effects of small changes to input parameters of these solutions in a sensitivity analysis study.

\subsection{Optimization Algorithm Background}

The first algorithm used for this study is Particle Swarm Optimization (PSO) that has been used in several studies that focus both on TFSC nanostructure optimization and also broader functions such as maximum power point tracking (MPPT) \cite{elkhamisy_comprehensive_2024,karim_optimizing_2024,sultan_particle_2024}. The effectiveness of the PSO algorithm as a general-purpose and computationally efficient optimization technique stems from its population-based search strategy, which is inspired by the collective flocking behavior observed in birds \cite{kennedy1995particle}. Within PSO, for each iteration the set of input parameters is organized in a parameter vector ($[R,t,c,p,z]$, for example) which is modeled as a particle with a position in space. Changes are applied to these parameters iteratively. The equations that summarize this process are as follows with Eq. \ref{eq:PSO1} and \ref{eq:PSO2} \cite{eberhart_new_optimizer}:

\begin{equation}
V_{i+1} = V_{i} + C_1 r_1 (P_{best} -P_i)+C_2 r_2 (G_{best} - P_i)
\label{eq:PSO1}
\end{equation}

\begin{equation}
P_{i+1} = P_i + V_{i+1}
\label{eq:PSO2}
\end{equation}

Here, $V_i$ represents the "velocity" and is calculated through adding a linear combination of the difference between $P_{best}$ (stored value of the best solution achieved for the particle) or $G_{best}$ (the best solution overall) and $P_i$ the current particle position. $C_1$ and $C_2$ are acceleration constants based on behavioral factors, while $r_1$ and $r_2$ are random numbers in the range of 0 to 1 (owing to the stochastic nature of the algorithm) \cite{eberhart_new_optimizer,sultan_particle_2024}. In this context, stochastic refers to the property where a random choice is made in the direction of search as the algorithm makes progress in fulfilling its objective iteratively \cite{Spall2003-xl}. The above equations applied to the particles are initialized at the beginning of a PSO process. To elaborate, similar to genetic algorithms, the PSO algorithm is initialized with a population; however, each particle receives its own velocity to traverse the solution space \cite{eberhart_new_optimizer}. A number of parallels can be drawn between the operation of PSO and genetic algorithms, and frequent use in obtaining optimal parameters for computational studies \cite{haque_effects_2023,zhou_study_2013}.

 A variant of genetic algorithm known as Differential Evolution (DE) has been used in this study. The availability of DE and Dual Annealing (DA) in the SciPy library has allowed the use of FDTD Solution's Python interoperability to use scripting to facilitate the automatic optimization process. DE is a robust, non-gradient method of searching the solution space. It is a stochastic algorithm and can rapidly converge on solutions \cite{storn_differential_1997}. However, the variation of DE used here utilizes a different mutation strategy. In the scripting, the specific mutation strategy used here is “DE/best/1” \cite{qiang_unified_2014,scipy_differential_evolution_nodate} and is different from a normal genetic algorithm’s method of mutation relying simply on a Gaussian distribution. For this study, the default mutation strategy of “DE/best/1” was used for all instances of optimization using DE.

Classical Simulated Annealing (CSA) is an algorithm based on the process of slowly cooling metal to achieve a more ordered lattice arrangement in a process known as annealing \cite{Van_Laarhoven1987-vj}. The movement of atoms in a metal lattice is governed by temperature. Temperature is an algorithm parameter of CSA which controls its stochastic property. Temperature decreases at a defined rate \cite{Van_Laarhoven1987-vj}  – usually exponential decay, over time.  As can be observed, the algorithms used in this study possess a stochastic property which provides the advantage of not getting trapped in local minima and maxima in most cases. This problem can be seen in the cases of gradient-based methods such as Steepest Descent \cite{karim_optimizing_2024,xin-she_yang_introduction_2019} applied to complex systems such as this. Different from the previous two, DA is not a population-based algorithm but takes steps according to each input and output. DA is a variation of the normal simulated annealing algorithm, being a combination of CSA and a variant algorithm called Fast Simulated Annealing (FSA) \cite{tsallis_generalized_1996}. This algorithm was selected for widespread known documentation of its use and behavior and has been utilized in FDTD simulations of TFSCs in previous studies \cite{kaya_extremely_2017}. DA is also a non-gradient method and can avoid getting trapped within local maxima or minima. PSO and DE being both population-based, have been given the same population size for each generation – 10. This can be ignored for DA which does not possess a factor such as population. Overall, the algorithms were run for a total of 1500-1800 simulations, allowing ample opportunity to converge on a potential maximum J\textsubscript{SC}. 

\section{Results and Discussion}

\subsection{Parameter Trend Analysis}

When testing the performance of the TFSC with respect to variations in different nanostructure parameters, it is important to examine how the J\textsubscript{SC} varies when multiple input parameters are varied simultaneously. Since there are five input nanostructure parameters as shown in Fig. \ref{fig:parameters} and one output performance parameter (J\textsubscript{SC} generated by the TFSC), plotting a graph to show the variation of J\textsubscript{SC} with respect to all five inputs would result in a 6-dimensional representation. A graph representing the variation of 5 inputs and 1 output cannot be easily understood. In order to determine an approximate relationship, this section explores contour plots plotted using simulation data collected by varying two different input parameters throughout the selected range shown in Table \ref{tab:parameter_constraints} and keeping the other three parameters constant. Thus, a 3-dimensional slice of the full 6-dimensional space can be graphically represented in this manner. This section aims to convey that analysis of the trends, though limited, can be helpful in understanding the broader relationship between the geometry and output. 

\begin{figure}[h]
    \centering
    \includegraphics[width=9cm]{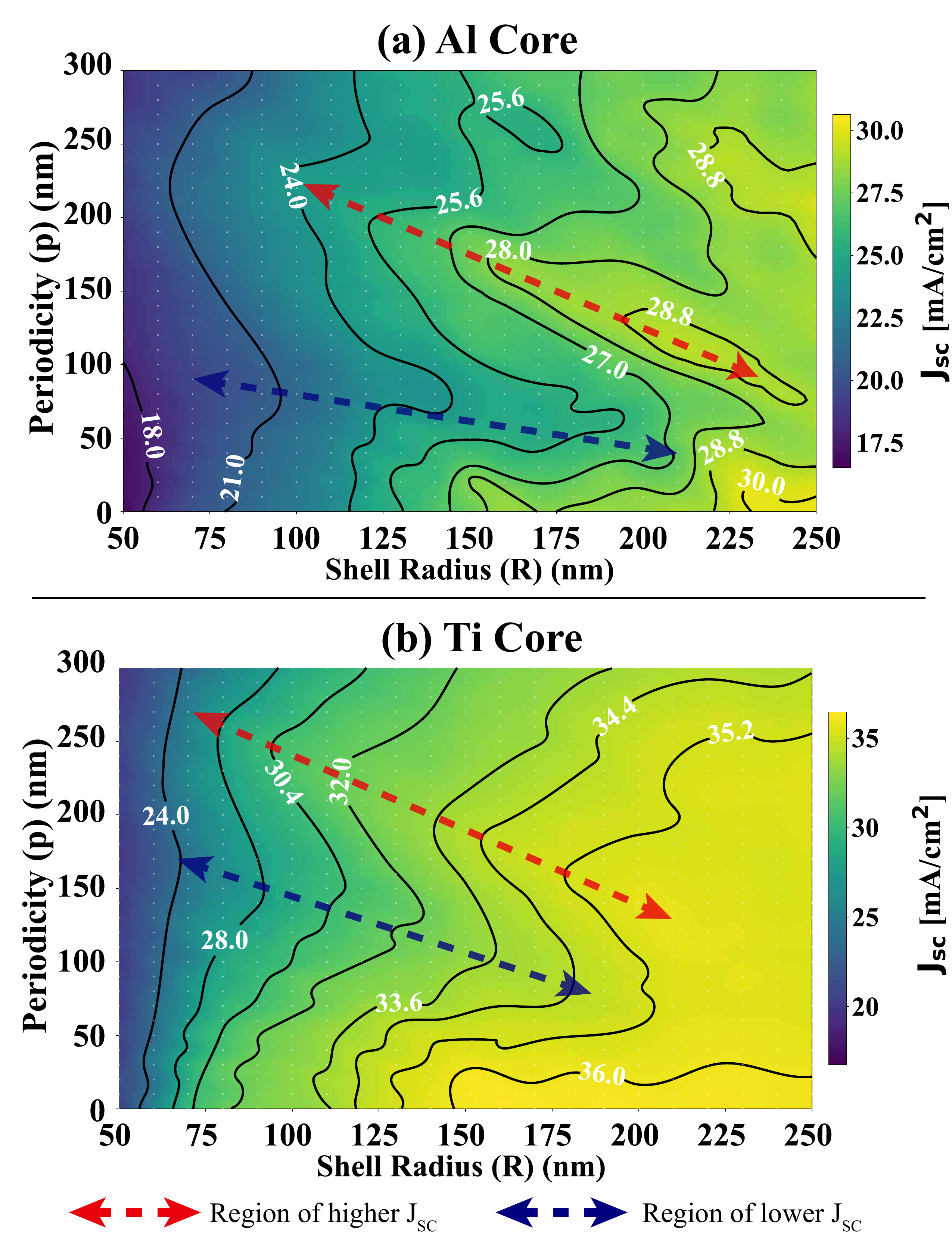}
    \caption{Contour plot detailing the J\textsubscript{SC} data obtained from parameter sweep of the periodicity $(p)$ and the radius $(R)$ of the triangular cross-section of the nanostructure. (a) Al core variation with respect to $R$ and $p$. (b) Ti core variation with respect to $R$ and $p$. The arrows annotating the plot highlight a specific pattern of combinations.}
    \label{fig:contour_pvR}
\end{figure}

Fig. \ref{fig:contour_pvR} shows the variation of J\textsubscript{SC} when the radius $(R)$ and periodicity $(p)$ of the nanostructure are varied while keeping shell thickness $(t)$, corner radius $(c)$, and vertical position $(z)$ constant. In Fig. \ref{fig:contour_pvR}, the x- and y-axes represent the parameters $R$ and $p$, respectively, while the color represents the J\textsubscript{SC} according to the corresponding scale bar. Data for this plot was generated through FDTD simulations where combinations of $R$ and $p$ were tested and the J\textsubscript{SC} was obtained. This process did not involve any optimization algorithm. The parameters $[c,t,z]$ are constant at $[50,15,500-R]$ in nanometers. Periodicity is the nearest distance between adjacent nanostructures. Fig. \ref{fig:contour_pvR} shows that generally an increasing value of $R$ leads to greater J\textsubscript{SC} for both core materials (Al and Ti in Fig. \ref{fig:contour_pvR}(a) and \ref{fig:contour_pvR}(b), respectively). Broadly, for higher values of $R$, lower values of $p$ can lead to higher J\textsubscript{SC}. As a result, it can be inferred that the best light trapping structures have a high value for $R$ and low values for $p$, thus pointing towards relatively larger nanostructures that are closer together. Trends such as these have also been observed in previous studies \cite{afnan_uzzaman_sheikh_investigating_2024,chowdhury_optimization_2024}. This can be true given that for both cores the brightest regions can be observed on the bottom right of the plot, where $R$ is high and $p$ is low. However, there are certain combinations that can be exceptions to this general rule such as the region around ($R$ = 225, $p$ = 100) for Fig. \ref{fig:contour_pvR}(a) and the region around ($R$ = 200, $p$ = 150) for Fig. \ref{fig:contour_pvR}(b). 

The double-headed arrows on the plots in Fig. \ref{fig:contour_pvR} roughly trace out the set of points that result in high or low J\textsubscript{SC}. However, the pattern is more prevalent in Fig. 5(a) than Fig. 5(b) because of the difference in core material. Instances of the pattern have been annotated on the plots in Fig. \ref{fig:contour_pvR} with double headed arrows. As established earlier, there is a general trend of higher values of $R$ and lower values of $p$ leading to higher J\textsubscript{SC}. However, on closer inspection of the plots, there are instances where lower $R$ and higher $p$ can provide high J\textsubscript{SC} values as well. The relation between J\textsubscript{SC} and input parameters can be complex, which is why optimization algorithms can be helpful in extracting the desired output across all variables. Variation of other parameters with $R$ are provided in Fig. S3 and Fig. S4 in the supplemental document.

\subsection{Optical Simulations (FDTD) Results}

Based on the J\textsubscript{SC} calculated with FDTD, sets of optimal parameters have been obtained for each core material and optimization algorithm. Once the optimal solutions/configurations are obtained, the drop in performance of the TFSC can be tested when subject to small random changes. Alongside this, the absorption of light for the improved TFSCs can be compared to a reference TFSC and the electric field intensity within the TFSC can be analyzed with near-field maps. 

\subsubsection{Optimization Algorithm Results and Discussion }

Tables \ref{tab:OptAl} and \ref{tab:OptTi} show the values of the input parameters obtained by the 3 optimization algorithms used, PSO, DE and DA for two different types of core materials, Al and Ti and their respective output J\textsubscript{SC} based on FDTD simulations. On average, each algorithm performed approximately 1600 simulations for the results shown in Tables \ref{tab:OptAl} and \ref{tab:OptTi}. As mentioned previously, the algorithms were supposed to find an optimum set of parameters as a requirement to fulfill the objective function as specified in Eq. \ref{eq:ObjFunc}. Owing to the stochastic (random) nature and varying strategies of the algorithms featured in this study, the three algorithms may not converge on same solution. The mean ($\mu$) and standard deviation ($\sigma$) are calculated in Tables \ref{tab:OptAl} and \ref{tab:OptTi}, for the set of optimized input parameters obtained from the three optimization algorithms (PSO, DE and DA) and J\textsubscript{SC} results. Additionally, the ratio ($\sigma / \mu$) $\times \ 100\%$   is the coefficient of variation (CV) and helps compare standard deviations relative to the mean.

\begin{table}[h]
\centering
\caption{Comparative analysis of the optimal nanostructure design parameters and corresponding J\textsubscript{SC} values obtained through optimization using three metaheuristic techniques—Particle Swarm Optimization (PSO), Differential Evolution (DE), and Dual Annealing (DA)—for the aluminum (Al) core nanostructure configuration.}
\label{tab:OptAl}
\begin{tabular}{|l|l|l|l|l|l|l|l|} \hline
Core                      & Algorithm      & $R$ (nm) & $t$ (nm) & $c$ (nm) & $p$ (nm) & $z$ (nm) & J\textsubscript{SC} (mA/cm\textsuperscript{2}) \\ \hline
\multirow{3}{*}{Al}       & PSO            & 250    & 10.0   & 59.4   & 27.5    & 0.00    & 33.6                           \\
                          & DE             & 249    & 10.1   & 25.8   & 162   & 0.00    & 32.9                           \\
                          & DA             & 247    & 11.0   & 13.0   & 168   & 0.00      & 32.8                           \\ \hline
\multicolumn{2}{|l|}{Mean $(\mu)$}               & 249    & 10.4   & 32.7   & 119    & 0.00    & 33.1                            \\ \hline
\multicolumn{2}{|l|}{Standard Deviation $(\sigma)$} & 1.53   & 0.55   & 24.0   & 79.4    & 0.00    & 0.44                            \\ \hline
\multicolumn{2}{|l|}{$CV = (\sigma/\mu) \times 100\ \%$}            & 0.61    & 5.31   & 73.2   & 66.7  & --   & 1.32 \\ \hline  
\end{tabular}
\end{table}

\begin{table}[h]
\centering
\caption{Comparative analysis of the optimal nanostructure design parameters and corresponding J\textsubscript{SC} values obtained through optimization using three metaheuristic algorithms—Particle Swarm Optimization (PSO), Differential Evolution (DE), and Dual Annealing (DA)—for the titanium (Ti) core nanostructure configuration.}
\label{tab:OptTi}
\begin{tabular}{|l|l|l|l|l|l|l|l|} \hline
Core                      & Algorithm      & $R$ (nm) & $t$ (nm) & $c$ (nm) & $p$ (nm) & $z$ (nm) & J\textsubscript{SC} (mA/cm\textsuperscript{2}) \\ \hline
\multirow{3}{*}{Ti}       & PSO            & 250    & 10.0   & 96.9   & 0.00    & 0.00    & 39.7                           \\
                          & DE             & 211    & 10.9   & 56.6   & 18.4   & 0.00    & 39.4                          \\
                          & DA             & 199    & 12.8   & 71.6   & 47.5   & 0.00      & 39.1                           \\ \hline
\multicolumn{2}{|l|}{Mean $(\mu)$}               & 220    & 11.2   & 75.0   & 22.0    & 0.00    & 39.4                            \\ \hline
\multicolumn{2}{|l|}{Standard Deviation $(\sigma)$} & 26.7   & 1.43   & 20.4   & 24.0    & 0.00    & 0.30                            \\ \hline
\multicolumn{2}{|l|}{$CV = (\sigma/\mu) \times 100\ \%$}            & 12.1    & 12.7   & 27.1   & 109  & --   & 0.76 \\ \hline  

\end{tabular}
\end{table}

The reference TFSC for comparison is a planar TFSC without a grating nanostructure and has an anti-reflective coating (ARC) without texture. The layer materials of the reference TFSC from top to bottom are SiO\textsubscript{2}, Si\textsubscript{3}N\textsubscript{4}, silicon, and Ag. For this reference TFSC, the layer thicknesses are 100 nm each for SiO\textsubscript{2} and Si\textsubscript{3}N\textsubscript{4}, 1000 nm for the silicon absorber layer and 100 nm for the Ag back reflector. FDTD simulation of the reference produces a J\textsubscript{SC} value of 17.6 mA/cm\textsuperscript{2}. Inspection of the results in Tables \ref{tab:OptAl} and \ref{tab:OptTi} reveal that the Al and Ti cores of the grating-like structure show a significant increase in J\textsubscript{SC} values compared to the reference TFSC. Comparing the mean J\textsubscript{SC} for both types of core (33.1 mA/cm\textsuperscript{2} for Al and 39.4 mA/cm\textsuperscript{2} for Ti), the Ti core structures help produce an output 19\% greater than that of the Al core structures. For both core materials, the standard deviation of J\textsubscript{SC} is low, but is larger for the input parameters as the algorithms use different search strategies. PSO finds the maximum possible values for J\textsubscript{SC} but DE and DA algorithms find very close J\textsubscript{SC} values at different input parameter values. It is to be noted that for parameters such as $c$ and $p$ in Table \ref{tab:OptAl} and \ref{tab:OptTi}, the CV values are very large. For example, the CV for $p$ in Table \ref{tab:OptTi} is greater than 100\% due to the wide range of variation. Most prominent of the results is the $z$ parameter which is 0 for all solutions. This can support the conjecture that the higher position of this grating-like structure aids in better scattering and helps trap reflected light off of the back reflector.  From these results, it can be inferred that both the Al and Ti cores can be a viable material for the nanostructures to produce TFSCs with enhanced optoelectronic performance. 

With respect to the optimized geometrical parameters, the tendency of $R$ and $t$ to converge toward the upper and lower bounds of the optimization space, respectively, can be attributed to the fact that increasing the effective volume of the metallic core enhances light scattering and plasmon-mediated energy transfer mechanisms \cite{karim_optimizing_2024}. Lastly, the $c$ and $p$ values for both types of core material in all algorithm solutions possess a high CV value, a detail that is relevant because high J\textsubscript{SC} can be achieved despite a high variation in these parameters. The values of $c$, which dictate the curvature of the edges, show no specific trends in maximizing J\textsubscript{SC}. Thus, indicating that strictly sharp edges are not essential for increased performance of TFSCs modified with grating-like nanostructures. This indicates that it might be possible to use cheaper (but less precise) fabrication methods to fabricate the core-shell nanostructures. The roundness of 2D periodic structures can be common due to elastic deformation phenomena, among other causes \cite{ok_continuous_2011}. Moreover, larger corner radius $(c)$ can lead to better scattering according to evidence provided by Fig. S3 in the supplemental document and in previous studies \cite{afnan_uzzaman_sheikh_investigating_2024}.

A valuable observation to note from the DE and DA solutions for Al is that they possess a large periodicity $(p)$ yet produce high J\textsubscript{SC} which is desirable due to the structures spaced further apart to assist in efficient light trapping. This means that less nanostructures need to be fabricated per unit area, reducing fabrication complexities and material usage requirements for each TFSC. The z position results can be an example of how optimization algorithms are also able to bring out certain solutions that can initially be counter-intuitive. A number of studies deal with gratings being fixed to the lower end or back-contact of the TFSC which can also provide substantial improvement to J\textsubscript{SC} and subsequently other electrical performance metrics \cite{heidarzadeh_performance_2019}.

\subsubsection{Optimal Solution Sensitivity Analysis}

\begin{figure}[b]
    \centering
    \includegraphics[width=12cm]{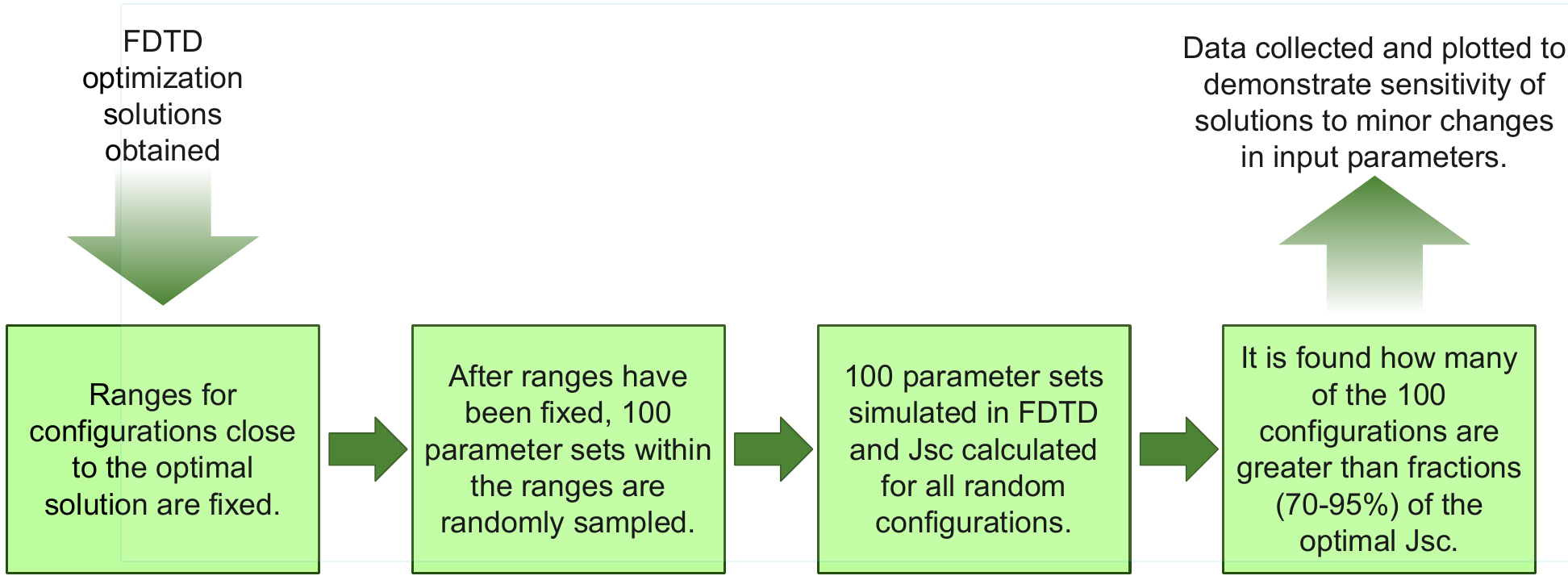}
    \caption{Process flow diagram illustrating the method for assessing sensitivity of optimal solutions brought about by PSO, DE and DA algorithms. The steps can be summarized as the obtainment of results, determining the ranges of variation, sampling parameters within those ranges and simulating, and counting how many samples cross a certain threshold.} 
    \label{fig:blockDiagram}
\end{figure}

Although this study is entirely simulation-based, it investigates a thin-film solar cell (TFSC) architecture that is potentially realizable for practical photovoltaic energy generation. The fabrication of such devices would necessitate advanced nanofabrication techniques and instrumentation capable of achieving nanometer-scale precision and dimensional control. However, for manufacturing purposes, less expensive techniques and equipment may be desirable from a cost perspective. Such low cost techniques may have the tradeoff of lower precision. Provided that there is such a restriction, it is useful to observe how minor changes to the parameters of the optimal solution affect the J\textsubscript{SC} generated by the modified TFSCs. For an optimal J\textsubscript{SC} value, there exists a corresponding specific combination of parameters. Thus, small changes to those parameters is expected to result in an output J\textsubscript{SC} less than the optimal value. Therefore, it is of interest to investigate how drastically the calculated J\textsubscript{SC} decreases for any small, random changes in the nanostructures structural parameters (i.e., input parameters).  An optimal solution which is more resistant to decrease in output J\textsubscript{SC} when subjected to these small changes in the input parameters is less sensitive and thus favorable as there can be less stringent standards for fabrication. Conversely, if small changes lead to sharper falls in the desired output, then a solution is more sensitive, and thus more precision is necessary to achieve similar results. The process proposed to measure the sensitivity of this multi-parameter system is illustrated in Fig. \ref{fig:blockDiagram} with a block diagram.

\begin{figure}[t]
    \centering
    \includegraphics[width=7cm]{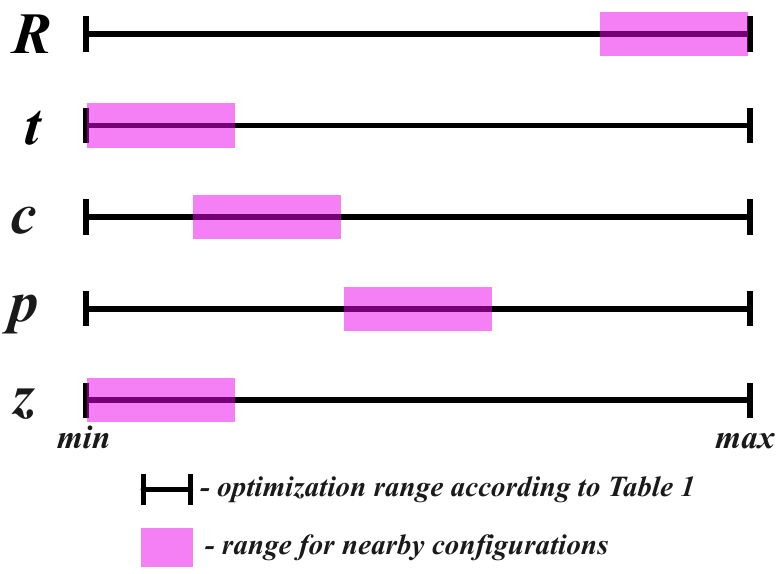}
    \caption{Diagram (not to scale) illustrating the range fixing for an example solution (DE-Al). The black line represents the aforementioned optimization range and the magenta shows the limits of the interval from which neighboring values can be selected from.} 
    \label{fig:rangeSens}
\end{figure}

Fig. \ref{fig:blockDiagram} presents a process flow diagram illustrating the methodological steps employed to evaluate the sensitivity of each optimized solution following the optimization procedure. Initially, the permissible bounds for parametric perturbations are defined based on the optimized parameter values obtained from the preceding optimization stage. Fig. \ref{fig:rangeSens} illustrates this for an example solution DE-Al, where the black line represents the full optimization range described in Table \ref{tab:parameter_constraints} and the magenta-colored sub-ranges highlight which values close to the optimal can be sampled from. To demonstrate, the ranges are arbitrarily $\pm 10\%$ of the optimization range around each optimal value, but are not allowed to go above or below the minimum or maximum value of the range shown in Table \ref{tab:parameter_constraints} (for example, $R$ cannot be greater than 250 nm). The ranges of variation for each optimal solution can be found in Table S1 of the supplemental document. 

The methods (detailed in the supplemental document) used to find the sub-ranges illustrated in Fig. \ref{fig:rangeSens} were done as a means of standardization. Depending on the variables such as fabrication machine precision or desired accuracy, the sub-ranges can be fixed accordingly. As mentioned in the second block shown in Fig. \ref{fig:blockDiagram}, after the ranges are fixed, a uniform random distribution is used to sample 100 sets of random parameters within these ranges. This method of sampling is similar to the process shown in \cite{kaya_extremely_2017} to produce data. These random sets of $[R,t,c,p,z]$ that are in the vicinity of the optimal result are then used to run FDTD simulations from which a value of J\textsubscript{SC} is calculated and stored. After this step is performed for each core metal (Al and Ti) and the three optimization techniques, it is then found how many of those random parameter combinations produce a J\textsubscript{SC} above threshold values (70\%-95\% of optimal J\textsubscript{SC}), as listed in Tables \ref{tab:OptAl} and \ref{tab:OptTi}. 

\begin{figure}[t]
    \centering
    \includegraphics[width=11.5cm]{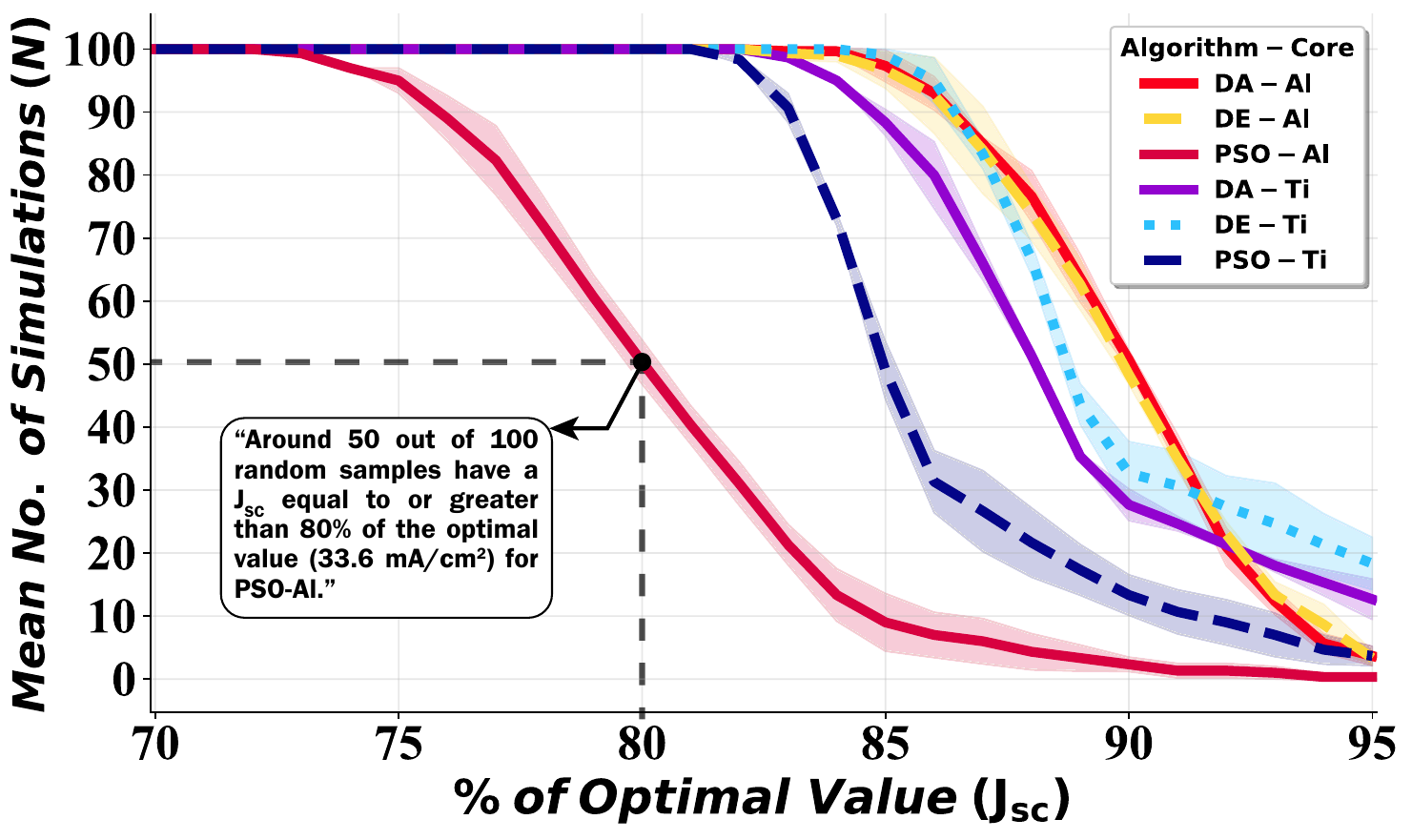}
    \caption{Sensitivity analysis of the optimized solutions presented as plots of the mean number of samples executed versus the percentage of the optimal J\textsubscript{SC}. As the generated J\textsubscript{SC} value from each run approaches the optimal J\textsubscript{SC} value for each solution, the number of samples among the 100 evaluated samples progressively decreases. The annotated point demonstrates the interpretation of an arbitrary point along the generated sensitivity curves}
    \label{fig:sensitivityAnalysis}
\end{figure}

As an example, the variation range for the DE-Al solution is produced, and 100 random sets of parameters are generated. Then FDTD simulations of those 100 parameter sets are simulated. As shown in Table \ref{tab:OptAl}, the J\textsubscript{SC} output of DE-Al is 32.9 mA/cm\textsuperscript{2} from the FDTD simulation. Then it is seen how many of those 100 simulations have an output equal to or greater than 70\%-95\% of 32.9 mA/cm\textsuperscript{2}. The same sequence of calculations is used for the other solutions as well, with respective optimal J\textsubscript{SC} values. For further rigor, the process described in Fig. \ref{fig:blockDiagram} is repeated twice more to find a mean and standard deviation of the number of simulations (greater than 70\%-95\% of the optimal J\textsubscript{SC}). These mean number of simulations are plotted in Fig. \ref{fig:sensitivityAnalysis} where the shaded region around each line is the standard deviation around the mean.

The final outcome of the process shown in Fig. \ref{fig:blockDiagram} are the results used to produce Fig. \ref{fig:sensitivityAnalysis}. It shows that for all optimal solutions, when the percentage of the optimal J\textsubscript{SC} on the horizontal axis increases, the number of parameter combinations that produce that J\textsubscript{SC}  decreases. Each line is a set of points that correspond to the number of samples that have J\textsubscript{SC} equal to or greater than a percentage of the optimal value. The shaded regions above and below the lines show the standard deviation around the mean number of samples that reach or surpass a threshold. Solutions that are less sensitive to decreases in performance when subjected to small changes are preferred. This lower sensitivity quality has been exhibited by three solutions most prominently as can be seen in Fig. \ref{fig:sensitivityAnalysis}, those being: DA-Al, DE-Al and DE-Ti. A notable observation is that samples for all of the solutions, when varied within the ranges specified in Table S1, produce J\textsubscript{SC} equal to or greater than 70\% of the optimal value. Another observation that can be made is that PSO-Al is the most sensitive, where only 50 samples out of 100 had J\textsubscript{SC} greater than or equal to 80\% of the optimal J\textsubscript{SC} (33.6 mA/cm\textsuperscript{2}). Whereas for the rest of the solutions, all 100 samples have J\textsubscript{SC} greater than 80\% of the optimal J\textsubscript{SC}. In general, solutions exhibiting lower sensitivity to small parameter perturbations are preferable, as enhanced robustness to fabrication-induced variations and nanofabrication tolerances may constitute an important design consideration alongside the maximization of device efficiency.  

\subsubsection{Absorption Spectra Analysis}

\begin{figure}[b]
    \centering
    \includegraphics[width=12cm]{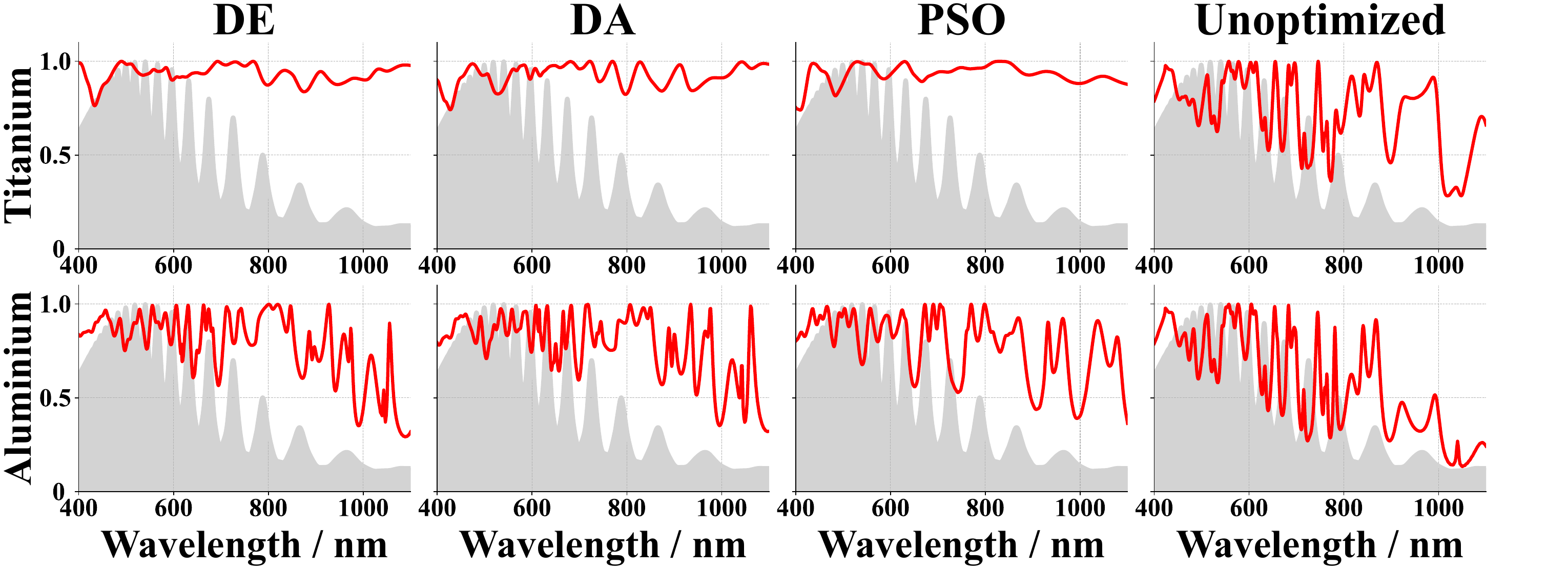}
    \caption{Absorbance spectra of incident light into the Si absorber layer of the TFSC. The red line represents the absorbance with the embedded structure present, and the gray area in the background is the absorbance of the reference (bare) TFSC.}
    \label{fig:abspec}
\end{figure}

The graphs depicted in Fig. \ref{fig:abspec} presents the absorption spectra of light within the silicon substrate as a function of wavelength. For comparison: the absorption spectrum for a planar silicon substrate with a non-textured ARC without nanostructures (the reference TFSC) is shaded in gray. The figure is structured such that the columns correspond to the optimization algorithm applied (DE, DA, and PSO) and the subsequent nanostructure configuration obtained from those algorithms and also the unoptimized data, while the rows represent the different core materials. For each case, the absorption profile of the optimized TFSC is plotted in red. In addition to the optimized structures, some data from unoptimized structures with physical parameters (in nm) of $[R, t, c, p, z] = [150, 40, 50, 150, 350]$, are presented as a reference as can be seen on the rightmost column in Fig. \ref{fig:abspec}. It can be seen that the spectra for the unoptimized structures very closely resemble the spectrum without the embedded structure. Across all optimization algorithms and core material types, the introduction of nanostructure significantly improves absorption across most of the spectral range. However, the TFSCs with Ti core nanostructures exhibit more stable absorption characteristics relative to Al core nanostructures, with enhanced broadband absorption efficiency, particularly in the 600–1100 nm wavelength range. It can be inferred that embedded structures assist in NIR photon energy conversion to generate electron-hole pairs. As a result, light trapping can be a contributor of lower energy photons being more readily absorbed into the substrate. It should also be noted that the photons according to $E = h \nu$ within the 400-1100 nm wavelength range possesses energy in the range of 1.12-3.10 eV \cite{Arons1965}.

\subsubsection{Near-Field Intensity Plots}

\begin{figure}[b]
    \centering
    \includegraphics[width=12cm]{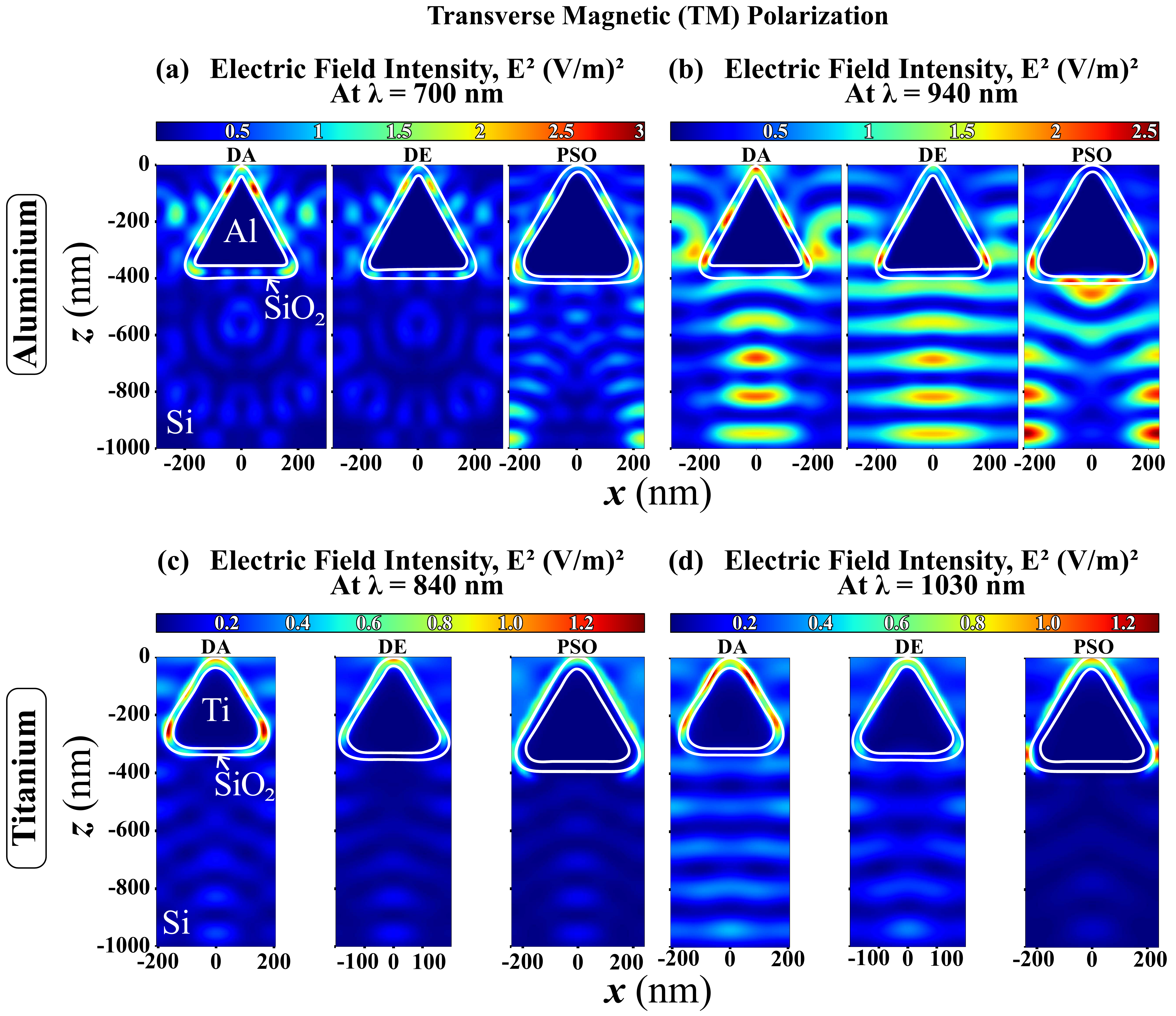}
    \caption{Near-field plots of Al and Ti embedded structure optimized by the three algorithms at wavelengths a) 700 nm, b) 940 nm for Al and c) 840 nm and d) 1030 nm for Ti. The near-field plots depict the electric field (E-field) energy distribution when illuminated under a TM incident light source.}
    \label{fig:NFTM}
\end{figure}

The near-field maps generated for the transverse magnetic (TM) polarization mode are shown for both core materials in Fig. \ref{fig:NFTM}. This figure shows the electric field squared intensity values within the silicon TFSC and embedded structure in the \textit{x-z} plane. For both types of core, two wavelengths have been sampled from the 400-1100 nm range. For the Al core, $\lambda$=700 nm and $\lambda$=940 nm have been sampled. For Ti core solutions, the wavelengths chosen for the near-field intensity plots were $\lambda$=840 nm and $\lambda$=1030 nm, respectively.

\begin{figure}[b]
    \centering
    \includegraphics[width=12cm]{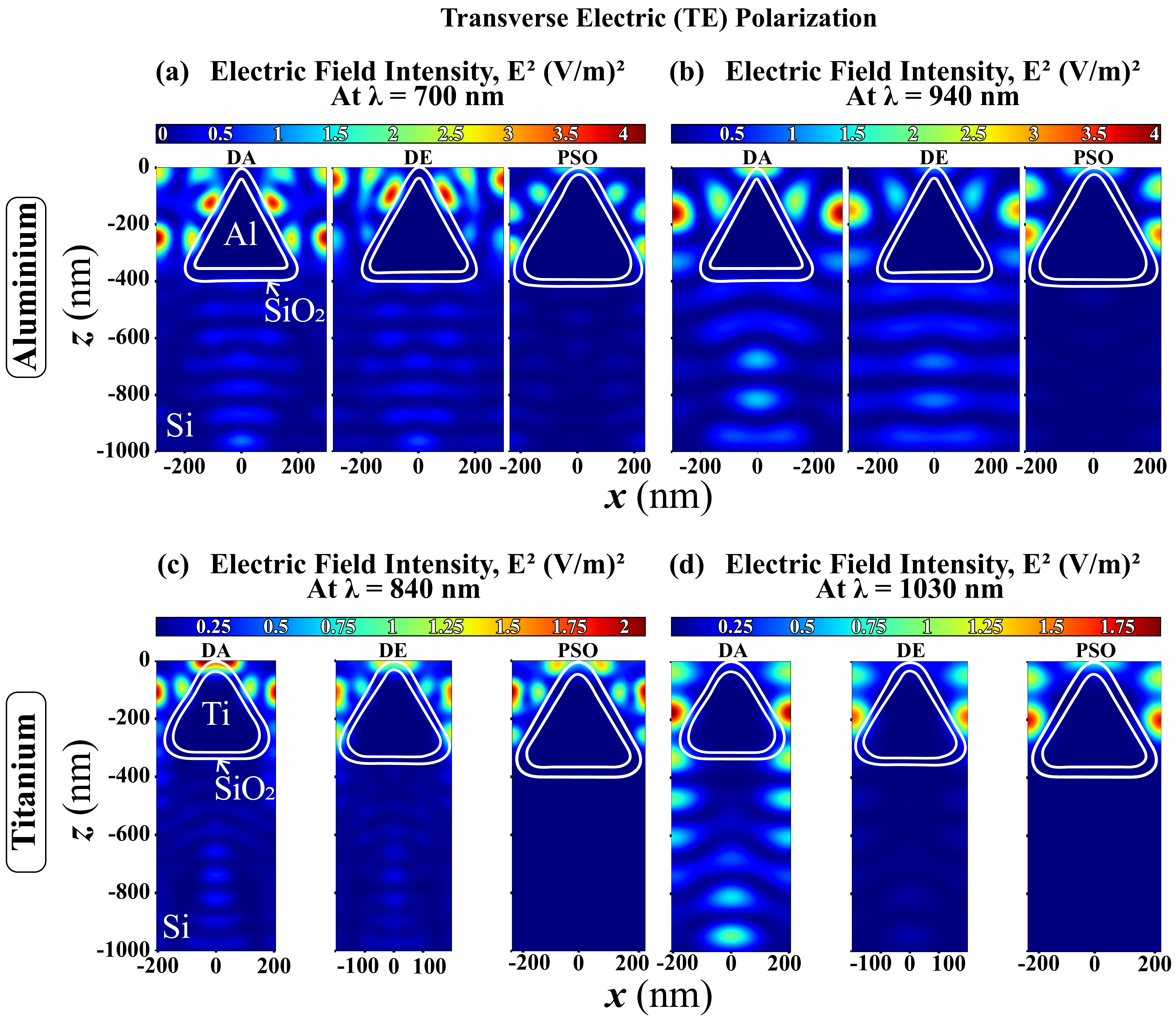}
    \caption{Near-field plots of Al and Ti embedded structure optimized by the three algorithms at wavelengths a) 700 nm, b) 940 nm for Al and c) 840 nm and d) 1030 nm for Ti. The near-field plots depict the electric field (E-field) energy distribution when illuminated under a TM incident light source.}
    \label{fig:NFTE}
\end{figure}

Generally, for all the solutions shown on Fig. \ref{fig:NFTM}, it can be seen that the electric fields intensify around or below the nanostructure, though the highest intensification is most consistently in the SiO\textsubscript{2} shell. The shell being an interface between the semiconductor and metal is a prime site for the propagation of surface plasmon polaritons (SPPs) \cite{zhu2017surface}. Moreover, SPPs throughout the length of nanostructures could be responsible for a portion of the energy transfer. LSPR can also contribute to energy transfer at these wavelengths as well, given that the high intensity E fields are always near the structure \cite{yu_effects_2017}. In Fig. \ref{fig:NFTM}(a), at 700 nm the PSO-Al solution also shows prominent intensification under the nanostructure apart from the shell compared to the DE-Al and DA-Al near-fields. For Fig. \ref{fig:NFTM}(b) at 940 nm the brightness of the intensification adjacent and below the nanostructure is much greater. This supports the idea of increased photocurrent generation from NIR photons resulting from the absorption increase compared to the reference TFSC in Fig. \ref{fig:abspec}. The intensity pattern seen for PSO-Al also supports the high J\textsubscript{SC} value found for the configuration from the optimization process. In Fig. \ref{fig:NFTM}(c) the intensification of the electric field is the most prominent within the shell except for the bottom face. The magnitude of $E^2$ below is not as intense as the values within the shell and to the sides of the nanostructure at 840 nm. For a higher wavelength of 1030 nm the ripple-like pattern below the structure is more obvious with the exception being PSO-Ti. The apex of the nanostructures seems to be the common region where the highest intensification can be located as can be seen in Fig. \ref{fig:NFTM}(c) and \ref{fig:NFTM}(d).

This section also analyzes the near-field map of the \textit{x-z} plane slice of the TFSC and the embedded nanostructure for TE polarized source, shown in Fig. \ref{fig:NFTE}. The patterns produced for both types of cores are different from the patterns that were observed in Fig. \ref{fig:NFTM}. The wavelengths sampled for this polarization is the exact same as the ones in Fig. \ref{fig:NFTM}. The most prominent patterns that can be observed for both cores irrespective of wavelength are the oblong spots of high intensification of the E-field above or to the sides of the triangular nanostructure. For the DE and DA solutions in Fig. \ref{fig:NFTE} ripple-like intensification patterns of the E-field can be witnessed as well. 

Regarding the differences in the near-field plots of Fig. \ref{fig:NFTM} and \ref{fig:NFTE}, it is understood that the interaction of surface plasmons in metallic nanostructures with incident electromagnetic radiation is strongly dependent on the polarization state of the incoming light. In the case of TM-polarized illumination, the electric field possesses a component normal to the metal–dielectric interface, enabling the excitation of collective oscillations of free electrons and efficient coupling to surface plasmon modes. Consequently, TM polarization typically produces stronger plasmonic resonance, enhanced near-field confinement, and increased optical absorption and scattering within the nanostructure. In contrast, TE-polarized light lacks a perpendicular electric-field component at the interface, thereby significantly limiting the excitation of conventional surface plasmon modes. As a result, the plasmonic response under TE polarization is generally weaker and is primarily governed by secondary effects such as geometric scattering, edge-induced field localization, or higher-order resonances in complex nanostructures. Therefore, the polarization-dependent behavior of metallic nanostructures arises fundamentally from the electromagnetic boundary conditions required for surface plasmon excitation at metal–dielectric interfaces \cite{nayeem_arefin_influence_2020,zhou_study_2013,li_research_surfacePlasmon_2025,afnan_uzzaman_sheikh_investigating_2024}.

\subsection{Electrical and Thermo-electrical Results Analysis}

According to the setup shown in Fig. \ref{fig:simSetup}(b) the electrical simulations of the optimal configurations for both Al and Ti cores were done. The generation rate data extracted from the FDTD simulations are imported to CHARGE to calculate the accurate values of J\textsubscript{SC} and the remaining performance metrics like V\textsubscript{OC}, FF, P\textsubscript{max} and $\eta$. Comparisons can be drawn between results where heat generation is not considered (at 300 K) and is considered (>300 K). Lastly, graphs of current  density (J) or output power per unit area (P) against voltage (V) are analyzed to compare results of optimal configurations against one another.

\subsubsection{Electrical Performance Metrics for Solar Cells}

The results obtained prior to this section were all computed solely using FDTD, which is mainly used for optical simulations. To identify the optimal device configuration, the short-circuit current density (J\textsubscript{SC}) was adopted as the optimization objective and figure of merit (FOM), with all candidate designs evaluated based on its maximization. This J\textsubscript{SC} is calculated using a scaled factor of the input power of the source and electron-hole pairs (e-h pairs) generated from light absorption, shown from Eq. \ref{eq:Pabs_raw}-\ref{eq:SCCD_Calc}. However, the current is limited by other factors that involve doping concentrations of the semiconductor, recombination of electron-hole pairs, resistance of the metal contacts, resistance of the absorbing material itself, etc. Hence, the current calculated with FDTD is higher than what is theoretically possible when the aforementioned factors are not accounted for. Additionally, the performance of a solar cell cannot be judged by J\textsubscript{SC} alone; there are other electrical parameters that define the quality of a solar cell. For better comparison, the J\textsubscript{SC} of the solar cell (after accounting for factors not included in FDTD), open-circuit voltage (V\textsubscript{OC}), fill factor (FF), maximum output power (P\textsubscript{max}) and efficiency ($\eta$) are calculated using the CHARGE solver based on the finite element method (FEM).  

A voltage sweep from 0 to 1.5 V is applied between the two metal contacts (cathode and anode) shown as top and bottom contacts in Fig. \ref{fig:simSetup}(b) to calculate the electrical parameters. The values of the relevant figures on which the CHARGE simulation is based have been provided in Table S2 of the supplemental document. The thicknesses are 15 nm and 25 nm for the p-doped region and n-doped region (shown in Fig. \ref{fig:simSetup}), respectively. The thicknesses and doping concentration values for these layers are taken from \cite{prentice2000computer}. A p-i-n junction TFSC is designed with an n-doped region at the top, intrinsic layer in the middle, and p-doped region at the bottom of the silicon.

\begin{figure}[h]
    \centering
    \includegraphics[width=10cm]{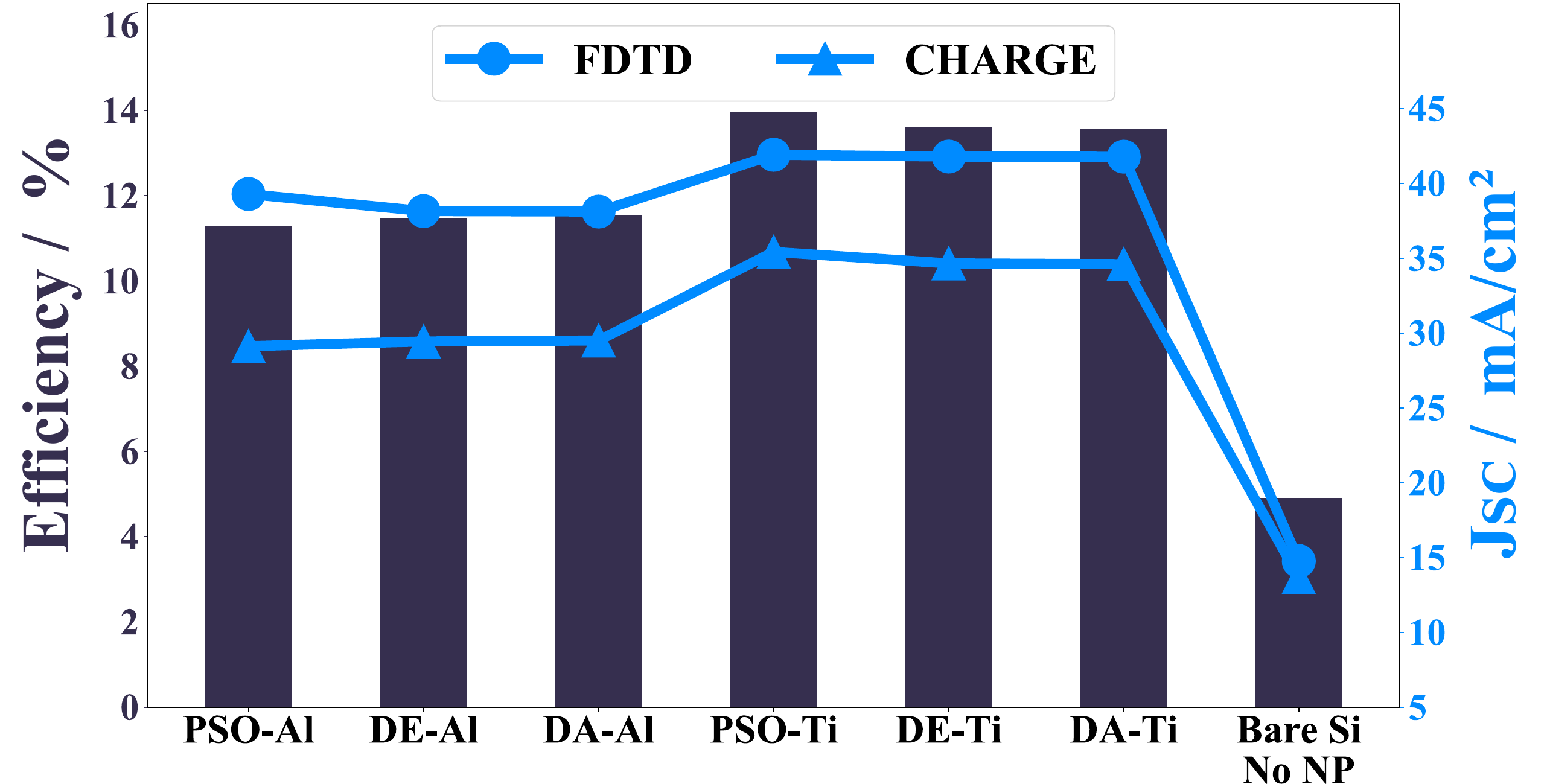}
    \caption{A direct comparison between the J\textsubscript{SC} obtained from FDTD calculations and CHARGE calculations. The difference between the two graphs resembles vertical translation in the vertical axis. A bar chart of efficiency is also present to visualize the variation}
    \label{fig:eff_Jsc}
\end{figure}

Table \ref{tab:elecResults} shows the electrical results of the six optimized configurations compared to the reference labeled "No NS". Here, "No NS" refers to "no nanostructure" - representing the reference TFSC where there is no embedded light trapping nanostructure and the ARC has no texture. As expected, the values of J\textsubscript{SC} computed in CHARGE decrease relative to the J\textsubscript{SC} of the same configuration computed using FDTD. The V\textsubscript{OC} and FF remain almost the same for all configurations. Fig. \ref{fig:eff_Jsc} shows the J\textsubscript{SC} calculated from CHARGE compared to FDTD. It shows the reduction in J\textsubscript{SC} while maintaining the same trend for both methods of calculation. This shows that optimization with FDTD simulations can be relied on for finding favorable parameters. Fig. \ref{fig:eff_Jsc} also shows the maximum efficiency calculated by all three optimization methods remains above 10\% and the maximum calculated efficiency is 13.93\% shown by the parameters determined by PSO using Ti core. For a solar cell, the efficiency is expected to increase when J\textsubscript{SC} increases. As expected, the change in efficiency follows a similar trend to J\textsubscript{SC}. Again, after optimization of the nanostructure, the one with the Ti core shows higher output than the one with Al. 

\begin{table}[b]
\centering
\caption{Electrical results for each of the Algorithm-Core solutions at 300 K.}
\label{tab:elecResults}
\begin{tabular}{|c|p{1.5cm}|p{1.5cm}|l|p{1.5cm}|l|l|}
\hline
Temperature (K) & Algorithm-Core & J\textsubscript{SC} (mA/cm\textsuperscript{2})    & V\textsubscript{OC}  (V)    & P\textsubscript{max} (mW/cm\textsuperscript{2})   & $\eta\  (\%)$     & FF       \\ \hline
\multirow{7}{*}{300}
& PSO Al         & 28.90 & 0.486  & 11.20 & 11.20 & 0.798 \\ \cline{2-7}
& DE Al          & 29.47 & 0.488 & 11.48 & 11.48 & 0.799 \\ \cline{2-7}
& DA Al          & 29.59 & 0.490 & 11.60 & 11.60 & 0.800 \\ \cline{2-7}
& PSO Ti         & 35.39 & 0.491 & 13.93 & 13.93 & 0.801 \\ \cline{2-7}
& DE Ti          & 34.73 & 0.491 & 13.64 & 13.64 & 0.800 \\ \cline{2-7}
& DA Ti          & 34.62 & 0.490 & 13.57 & 13.57 & 0.800 \\ \cline{2-7} 
& No NS  & 13.69 & 0.464 & 4.91 & 4.91 & 0.773 \\ \hline
\end{tabular}
\end{table}

The simulations were done at a temperature of 300 K. However, as a solar cell is exposed to light, many of the absorbed high energy photons generate heat after the creation of electron-hole pairs. Additionally, the movement of charges within the semiconductor increases the temperature of the TFSC as well. At a certain temperature the heat gained by the solar cell becomes equal to the heat lost to the environment. When this condition is reached, the temperature of the solar cell does not change further. This average internal temperature of the TFSC at this stage is known as the steady-state temperature \cite{planar_si_sc}. At steady-state temperature, the electrical output parameters of the solar cell can vary. As such, another set of CHARGE simulations were performed to take this heating effect into account. The corresponding steady state temperatures and electrical performance parameters are provided in Table S3 in the supplemental document. Observing the values in Table S3, the maximum increase in temperature is 20 K for the DA-Al core and the minimum increase is 8 K for PSO-Ti core. Except for these two values, temperature increase generally stays within the range of 14 to 17 K. The general trend found is that an increase in temperature leads to a drop in performance in the solar cell, the exceptions being PSO-Al and PSO-Ti. A possible reason for this could be the low values for $p$ for these configurations. Despite the fact that conductivity in semiconductors usually increases with temperature, the TFSC for this study is not one continuous volume. The structure is not uniform in the volume of the silicon due to the embedded nanostructure. and only the average temperature is being calculated. As seen in the near-fields in Fig. \ref{fig:NFTM} and \ref{fig:NFTE}, intensification of the electric field often occurs in the dielectric shell. So, localized effects of temperature are not being considered.  Changes in the values of electric performance parameters are listed in Table S4 of the supplemental document. The drop in J\textsubscript{SC} for DE-Al and DA-Al is 0.03 and 0.09 mA/cm\textsuperscript{2} respectively. For DE-Ti and DA-Ti the drop is 0.08 and 0.01 mA/cm\textsuperscript{2} respectively. This drop in J\textsubscript{SC} is negligible and does not significantly affect the performance of the solar cell. However, for the PSO-Al and PSO-Ti configurations, the J\textsubscript{SC} increased by 0.22 and 0.03 mA/cm\textsuperscript{2} respectively. An increase in 0.03 mA/cm\textsuperscript{2} in the case of PSO-Ti configuration is still very marginal. An increase of 0.22 mA/cm\textsuperscript{2} for the PSO-Al configuration is greater relative to the other configurations, albeit a change from 28.90 to 29.11 mA/cm\textsuperscript{2} cannot be considered significant. Generally for the results obtained, the magnitudes of the changes are very small, thus the change in device performance levels in this context are not very significant. Even though the maximum increase in temperature among the six different configurations is 20 K, i.e., a rise from 300 K (27 $\degree$C) to 320 K (47 $\degree$C), the change in the performance of the TFSC remains relatively stable. Hence the simulation results suggest that the modeled TFSC exhibits relatively low temperature sensitivity, indicating improved thermal stability under varying operating conditions.

\subsubsection{Current Density (J) and Power per unit Area (P) Against Voltage (V) Graphs}

\begin{figure}[h]
    \centering
    \includegraphics[width=13cm]{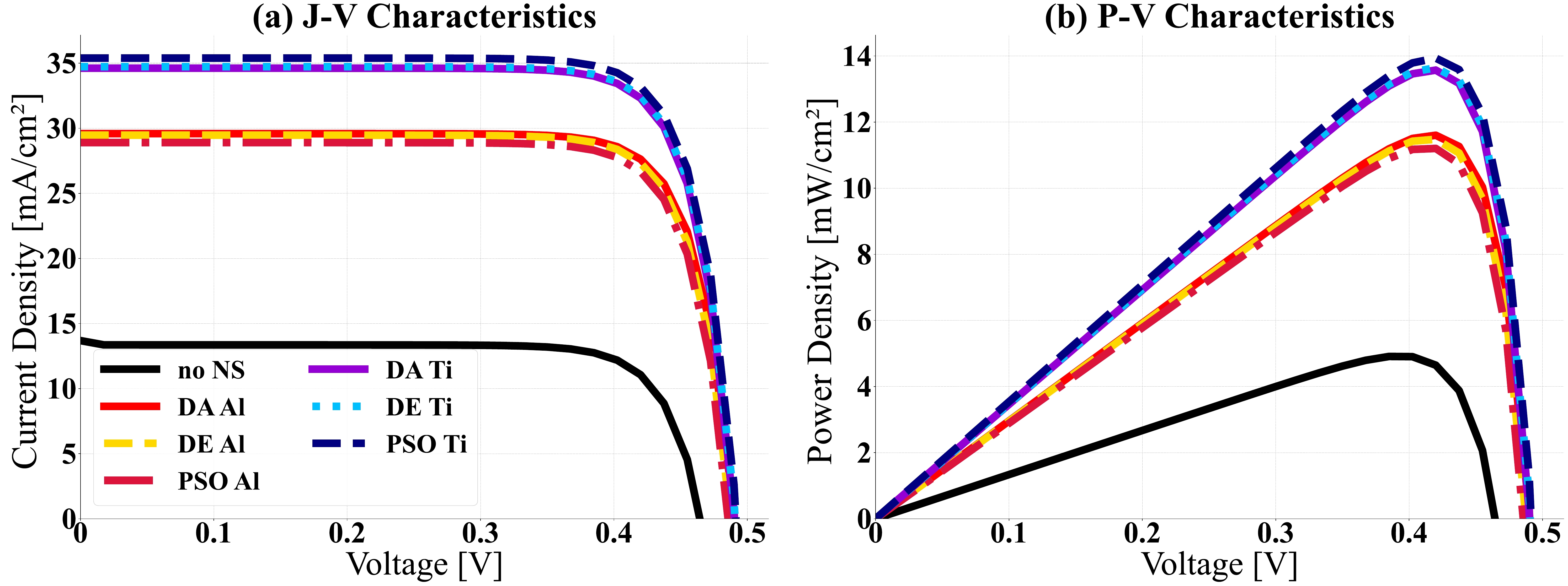}
    \caption{Plots produced from CHARGE simulations. In (a) the current density (J) is plotted against voltage (V) and in (b) the power per unit area (P) against voltage (V) is plotted.}
    \label{fig:JVPV}
\end{figure}

Fig. \ref{fig:JVPV} presents the J-V and P-V characteristics of the TFSCs with the configurations optimized using the algorithms along with the aforementioned reference TFSC. The J-V curve in the Fig. \ref{fig:JVPV}(a) refers to current density against voltage and the P-V curve in Fig. \ref{fig:JVPV}(b) is the power density against voltage.

Ideally, the shape of the J-V curve should be completely flat until the voltage reaches the V\textsubscript{OC}, and a sharp drop in J\textsubscript{SC} when the voltage reaches V\textsubscript{OC}. The ideal shape indicates that the solar cell can output the same current over the entire voltage range of the device. It also implies zero series internal resistance of the solar cell. Practically, the shape is as shown in Fig. \ref{fig:JVPV}(a) where the slightly non-flat curve indicates low but non-zero internal resistance of the solar cell caused by reasons such as the recombination of the electron-hole pairs, contact resistance, the limited conductivity of the silicon, etc. 

For all six configurations, the gradient of the J-V curve remained almost 0 up to 0.35 V. The curves slowly decrease from the point where $J \approx J_{SC}$ to 0 mA/cm\textsuperscript{2} as the voltage rises from 0.35 V to 0.49 V, which is the open-circuit voltage (V\textsubscript{OC}). The power per unit area against voltage (P-V) curve is plotted by computing the product of J and V at every discrete voltage point. The J-V curve shows that as V increases, J decreases, which means that while computing $(J \times V)$ there is a specific point in the P-V curve where the value of the product is maximum. The maximum point on the P-V curve in Fig. \ref{fig:JVPV}(b) refers to P\textsubscript{max} which is the maximum power delivered by a single solar cell. This point where the power is maximum can be called the maximum power point (MPP). At MPP, the cell outputs a current density of J\textsubscript{MPP} at a voltage of V\textsubscript{MPP}. P\textsubscript{max} is the maximum power that the solar cell can deliver if a load attached to the cell can draw a current of J\textsubscript{MPP} that will cause the cell to output a voltage equal to V\textsubscript{MPP} \cite{kharb_modeling_2014}. The P\textsubscript{max} for the TFSCs with Ti core nanostructures on average is 13.7 mW/cm\textsuperscript{2}, and is 20\% greater than those with the TFSCs with Al core structures with an average P\textsubscript{max} of 11.4 mW/cm\textsuperscript{2}.

\section{Potential Fabrication Process}

\begin{figure}[h]
\centering\includegraphics[width=9cm]{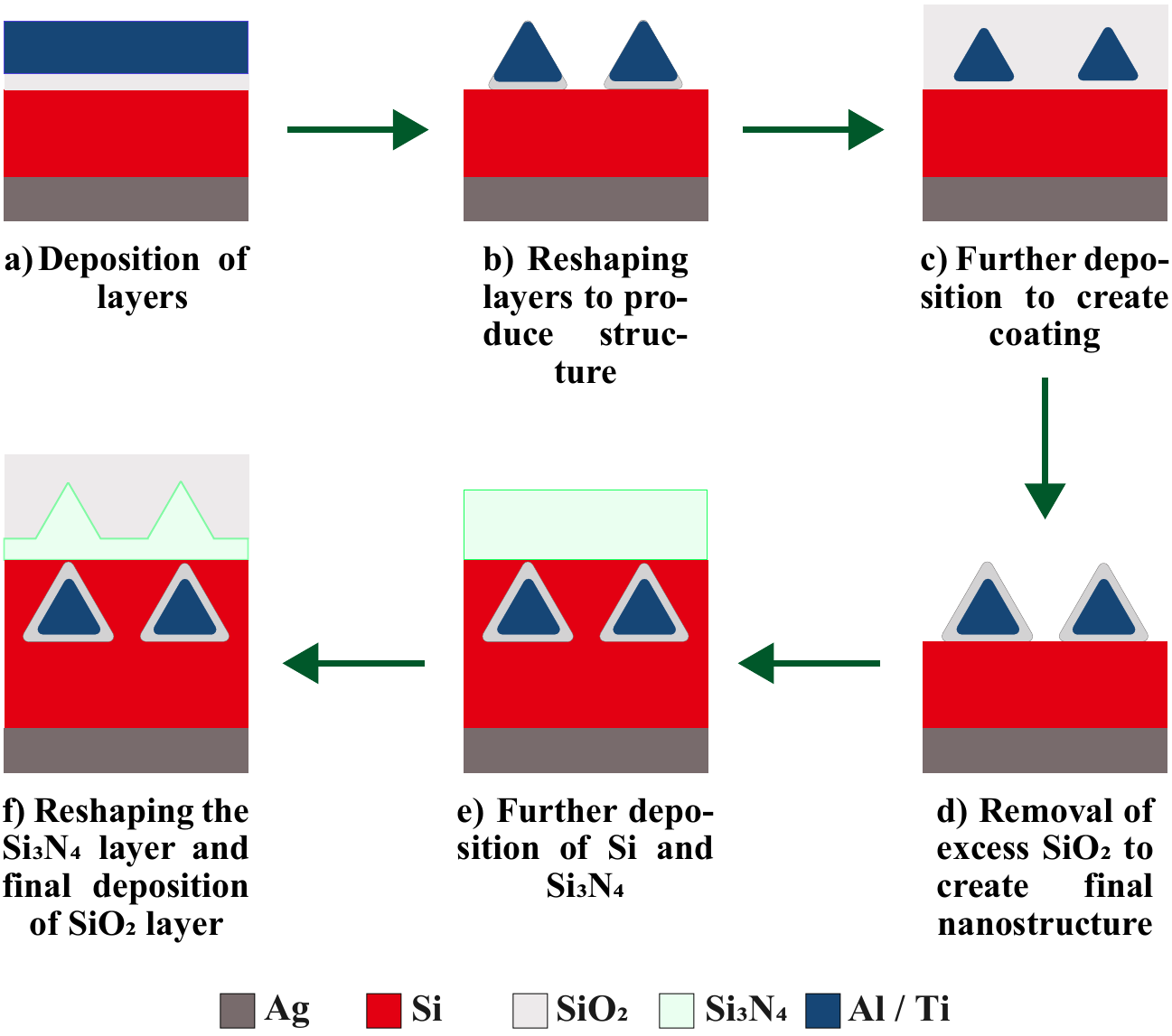}
\caption{A proposed sequence of fabrication for the TFSC design in this study. (a) Deposition of Ag, Si, Al and SiO\textsubscript{2} layers with magnetron sputtering, PECVD, and ALD where appropriate. (b) Use of IBL, EBL or IL to produce the nanostructure core. (c) SiO\textsubscript{2} deposition with PECVD or ALD. (d) Removal of excess SiO\textsubscript{2} with IBL. (e) Si\textsubscript{3}N\textsubscript{4} layer deposited with plasma nitriding. (f) NIL or IBL create the Si\textsubscript{3}N\textsubscript{4} texture, SiO\textsubscript{2} is deposited to complete the ARC.}
\label{fig:fabrication_updated}
\end{figure}

After proper review of literature relating to fabrication of complex TFSCs and other layered devices, this section suggests possible steps to fabricate the TFSC design in this study. The general course of this process is summarized within Fig. \ref{fig:fabrication_updated} through alternation of deposition and lithography techniques. The layer-wise deposition techniques vary by material. To begin with, the layers must be deposited in the following order, bottom to top as can be seen in Fig. \ref{fig:fabrication_updated}(a). The Ag back-reflector layer can be produced with magnetron sputtering \cite{li_aluminum_2020, eze_optimum_2021}, and the silicon layer can be produced with plasma-enhanced chemical vapor deposition (PECVD) \cite{haug_light_2015}. On top of the silicon absorber layer, a layer of SiO\textsubscript{2} is added, on which a layer of Al or Ti is deposited further. SiO\textsubscript{2} can be deposited with techniques such as atomic layer deposition (ALD) or PECVD \cite{arl_sio2_2020,ping_pecvd_2011} and the Al film can be deposited with the assistance of magnetron sputtering. In Fig. \ref{fig:fabrication_updated}(b), the Al or Ti layers are etched with focused ion beam (FIB) or ion beam lithography (IBL) \cite{seniutinas_nano-proximity_2016}. A possible alternative to IBL would be electron beam lithography (EBL) or interference lithography (IL) which have been documented to fabricate periodic Al nanostructures \cite{martin_fabrication_2015}. Fig. \ref{fig:fabrication_updated}(c) shows further deposition of SiO\textsubscript{2} which can be done with PECVD or ALD as mentioned above, and this is done to form the rest of the coating of the embedded structure. A study by Poitras et al. documents a method of precise layer control with a system that can alternate between etching and deposition. The study also states that IBL can be used for etching SiO\textsubscript{2} \cite{poitras_ion-beam_2003} as required in Fig. \ref{fig:fabrication_updated}(d). A Si\textsubscript{3}N\textsubscript{4} layer can possibly be deposited in Fig. \ref{fig:fabrication_updated}(e) with PECVD or plasma nitriding \cite{liu_study_Si3n4_TFs_2013,chen_growth_2020}. Lastly, as shown in Fig. \ref{fig:fabrication_updated}(f) the Si\textsubscript{3}N\textsubscript{4} layer can be textured with nanoimprint lithography (NIL) or IBL \cite{yoshinaga_low_2017, Mitrofanov2017} and the ARC can be completed through the deposition of SiO\textsubscript{2}.

\section{Conclusion}

This study presents a computational investigation of a novel triangular core-shell grating-inspired light trapping nanostructure embedded within the photoactive layer of silicon thin-film solar cells (TFSCs). The optical and optoelectronic behavior of the modified Si TFSC architecture was systematically analyzed using the finite-difference time-domain (FDTD) method, with device parameters optimized to maximize photovoltaic performance levels. The optical results were further analyzed using CHARGE simulations. A wide range of designs are possible within the specified parameter ranges. The objective of the study is to extract a combination of parameters that favors improved optoelectronic performance. As a result, three optimization algorithms were applied to generate the optimal results. The goal of the algorithms was to maximize the short-circuit current density (J\textsubscript{SC}) which in turn would require more electron-hole pairs to be produced. The algorithms used were: Particle Swarm Optimization (PSO), Differential Evolution (DE) and Dual Annealing (DA). The nanostructure core was implemented using aluminum (Al) and titanium (Ti), respectively, to assess their impact on device performance. For each core, three algorithms were tested and six optimized solutions were obtained as a result. The range of J\textsubscript{SC} covered varied from 28.90 to 35.39 mA/cm\textsuperscript{2} compared to the reference (i.e., a TFSC with no nanostructures, which has a J\textsubscript{SC} of 13.69 mA/cm\textsuperscript{2}). It is observed that there is an improvement by a factor of around 2.1 to 2.5 (i.e., approximately a 200\% improvement in performance). This factor of improvement can also be witnessed in terms of efficiency. An unmodified TFSC has an efficiency of 4.9\% and a TFSC with optimized light trapping structure has an efficiency of 12.6\% on average for all solutions (i.e., more than 200\% enhancement). The effects of temperature variations were also studied, and it was seen that the simulated TFSC performance parameters are relatively stable to internal (i.e., absorbing layer) temperature increases at an ambient temperature of 300 K. It was further demonstrated that, beyond identifying an optimal set of input parameters and the corresponding output response, additional analyses of the simulated physical system are necessary to reach a more nuanced and practically relevant assessment. An example of an analysis assisting in reaching a more robust understanding was testing the sensitivity of each solution to small perturbations. Future studies could focus on more robust designs for TFSCs, considering a larger number of variables. Other algorithms and machine learning techniques could be implemented to create useful surrogate models. Lastly, other performance metrics can be used as figures of merit for optimization. For example, minimization of reflectivity or maximization of QE, to gain further insight into what factors are most important to consider to enhance TFSC performance.

\begin{backmatter}

\bmsection{Acknowledgment} The authors wish to express gratitude to Independent University, Bangladesh (IUB) for funding this research and for providing additional logistical support.

\bmsection{Funding} Independent University, Bangladesh (IUB) (Sponsored Research Project No: 2025-SETS:R20).

\bmsection{Disclosure} The authors declare no conflict of interest.

\bmsection{Data Availability} Data underlying the results presented in this paper are not publicly available at this time but may be obtained from the authors upon reasonable request.

\bmsection{Supplemental document} See Supplement 1 for supporting content.

\end{backmatter}


\bibliography{journalBib}

\begin{thebibliography}{10}
\newcommand{\enquote}[1]{``#1''}

\bibitem{kannan_solar_2016}
N.~Kannan and D.~Vakeesan, \enquote{Solar energy for future world: - {A}
  review,} {\protect\JournalTitle{Renewable and Sustainable Energy Reviews}}
  \textbf{62}, 1092--1105 (2016).

\bibitem{maka_solar_2022}
A.~O.~M. Maka and J.~M. Alabid, \enquote{Solar energy technology and its roles
  in sustainable development,} {\protect\JournalTitle{Clean Energy}}
  \textbf{6}, 476--483 (2022).

\bibitem{elkhamisy_comprehensive_2024}
K.~ElKhamisy, H.~Abdelhamid, E.-S.~M. El-Rabaie, and N.~Abdel-Salam, \enquote{A
  {Comprehensive} {Survey} of {Silicon} {Thin}-film {Solar} {Cell}:
  {Challenges} and {Novel} {Trends},} {\protect\JournalTitle{Plasmonics}}
  \textbf{19}, 1--20 (2024).

\bibitem{li_flexible_2024}
Y.~Li, X.~Ru, M.~Yang, \emph{et~al.}, \enquote{Flexible silicon solar cells
  with high power-to-weight ratios,} {\protect\JournalTitle{Nature}}
  \textbf{626}, 105--110 (2024). Publisher: Nature Publishing Group.

\bibitem{panagiotopoulos_critical_2023}
A.~Panagiotopoulos, T.~Maksudov, G.~Kakavelakis, \emph{et~al.}, \enquote{A
  critical perspective for emerging ultra-thin solar cells with ultra-high
  power-per-weight outputs,} {\protect\JournalTitle{Applied Physics Reviews}}
  \textbf{10}, 041303 (2023).

\bibitem{otte_flexible_2006}
K.~Otte, L.~Makhova, A.~Braun, and I.~Konovalov, \enquote{Flexible
  {Cu}({In},{Ga}){Se2} thin-film solar cells for space application,}
  {\protect\JournalTitle{Thin Solid Films}} \textbf{511-512}, 613--622 (2006).

\bibitem{poortmans_thin_2006}
J.~Poortmans and V.~Arkhipov, eds., \emph{Thin {Film} {Solar} {Cells}:
  {Fabrication}, {Characterization} and {Applications}} (Wiley, 2006), 1st ed.

\bibitem{brongersma_light_2014}
M.~L. Brongersma, Y.~Cui, and S.~Fan, \enquote{Light management for
  photovoltaics using high-index nanostructures,} {\protect\JournalTitle{Nature
  Materials}} \textbf{13}, 451--460 (2014). Publisher: Nature Publishing Group.

\bibitem{haug_light_2015}
F.-J. Haug and C.~Ballif, \enquote{Light management in thin film silicon solar
  cells,} {\protect\JournalTitle{Energy \& Environmental Science}} \textbf{8},
  824--837 (2015).

\bibitem{nayeem_arefin_influence_2020}
S.~M. Nayeem~Arefin, J.~Hasan, S.~Islam, \emph{et~al.}, \enquote{Influence of
  {Particle} {Shape} on the {Ability} of {Plasmonic} {Metal} {Core}-{Silica}
  {Shell} {Nanoparticles} {Embedded} within the {Absorbing} {Layer} to
  {Enhance} the {Opto}-electronic {Performance} of {Thin}-{Film} {Solar}
  {Cells},} in \emph{2020 2nd {International} {Conference} on {Advanced}
  {Information} and {Communication} {Technology} ({ICAICT}),}  (IEEE, Dhaka,
  Bangladesh, 2020), pp. 416--421.

\bibitem{huang_thin-film_2022}
Z.~Huang and B.~Wang, \enquote{Thin-{Film} {Solar} {Cells} by {Silicon}-{Based}
  {Nano}-{Pyramid} {Arrays},} {\protect\JournalTitle{Advanced Theory and
  Simulations}} \textbf{5}, 2100586 (2022). \_eprint:
  https://onlinelibrary.wiley.com/doi/pdf/10.1002/adts.202100586.

\bibitem{tsai_handbook_2023}
C.-W.~C. TSAI, \emph{{HANDBOOK} {OF} {METAHEURISTIC} {ALGORITHMS}: from
  fundamental theories to advanced applications}, Uncertainty, computational
  techniques, and decision intelligence (ELSEVIER ACADEMIC PRESS, S.l., 2023).

\bibitem{Li2019}
F.~Li, X.~Peng, Z.~Wang, \emph{et~al.}, \enquote{Machine learning
  (ml)‐assisted design and fabrication for solar cells,}
  {\protect\JournalTitle{ENERGY \& ENVIRONMENTAL MATERIALS}} \textbf{2},
  280–291 (2019).

\bibitem{karim_optimizing_2024}
A.~Karim, A.~F. M. A.~U. Sheikh, S.~A. Chowdhury, and M.~H. Chowdhury,
  \enquote{Optimizing the {Parameters} of {Core}-{Shell} {Nanoparticles} with
  {Different} {Algorithms} to {Enhance} {Performance} of {Thin}-{Film} {Solar}
  {Cells},} in \emph{2024 {International} {Conference} on {Advances} in
  {Computing}, {Communication}, {Electrical}, and {Smart} {Systems}
  ({iCACCESS}),}  (2024), pp. 1--6.

\bibitem{zhou_study_2013}
S.~Zhou, X.~Huang, Q.~Li, and Y.~M. Xie, \enquote{A study of shape optimization
  on the metallic nanoparticles for thin-film solar cells,}
  {\protect\JournalTitle{Nanoscale Research Letters}} \textbf{8}, 447 (2013).

\bibitem{haque_effects_2023}
A.~J. Haque, A.~A. Suny, R.~B. Sultan, \emph{et~al.}, \enquote{Effects of
  “defective” plasmonic metal nanoparticle arrays on the opto-electronic
  performance of thin-film solar cells: computational study,}
  {\protect\JournalTitle{Applied Optics}} \textbf{62}, 3028--3041 (2023).
  Publisher: Optica Publishing Group.

\bibitem{muhammad_optimization_2018}
M.~H. Muhammad, K.~R. Mahmoud, M.~F.~O. Hameed, and S.~S.~A. Obayya,
  \enquote{Optimization of highly efficient random grating thin-film solar cell
  using modified gravitational search algorithm and particle swarm optimization
  algorithm,} {\protect\JournalTitle{Journal of Nanophotonics}} \textbf{12}, 1
  (2018).

\bibitem{hasan_use_2021}
S.~N. Hasan, A.~J. Haque, T.~A. Khan, and M.~H. Chowdhury, \enquote{Use of
  {Aluminum}-{Silica} {Core}-{Shell} {Plasmonic} {Nanoparticles} to {Enhance}
  the {Opto}-{Electronic} {Performance} of {Thin}-{Film} {Solar} {Cells},} in
  \emph{2021 6th {International} {Conference} on {Development} in {Renewable}
  {Energy} {Technology} ({ICDRET}),}  (IEEE, Dhaka, Bangladesh, 2021), pp.
  1--6.

\bibitem{Maier2007-xg}
S.~Maier, \emph{Plasmonics: Fundamentals and Applications} (Springer, New York,
  NY, 2007).

\bibitem{sultan_particle_2024}
R.~B. Sultan, A.~A. Suny, M.~H. Hossain, \emph{et~al.}, \enquote{Particle swarm
  optimization for performance enhancement of cadmium telluride thin film solar
  cells by embedded nano-grating structures with a plasmonic metal coating
  layer,} {\protect\JournalTitle{Heliyon}} \textbf{10}, e38775 (2024).
  Publisher: Elsevier BV.

\bibitem{elsheikh_review_2019}
A.~H. Elsheikh and M.~Abd~Elaziz, \enquote{Review on applications of particle
  swarm optimization in solar energy systems,}
  {\protect\JournalTitle{International Journal of Environmental Science and
  Technology}} \textbf{16}, 1159--1170 (2019).

\bibitem{kaya_extremely_2017}
M.~Kaya and S.~Hajimirza, \enquote{Extremely {Efficient} {Design} of {Organic}
  {Thin} {Film} {Solar} {Cells} via {Learning}-{Based} {Optimization},}
  {\protect\JournalTitle{Energies}} \textbf{10}, 1981 (2017). Number: 12
  Publisher: Multidisciplinary Digital Publishing Institute.

\bibitem{hara_optimum_1992}
K.~Hara, T.~Iwamoto, and K.~Kyuma, \enquote{Optimum design method of
  optoelectronic devices using simulated annealing,}
  {\protect\JournalTitle{IEEE Photonics Technology Letters}} \textbf{4},
  1360--1362 (1992).

\bibitem{ansys_lumerical_suite}
\enquote{Ansys {Lumerical} {FDTD} {\textbar} {Simulation} for {Photonic}
  {Components},} .

\bibitem{planar_si_sc}
\enquote{Planar silicon solar cell --- optics.ansys.com,}
  \url{https://optics.ansys.com/hc/en-us/articles/360042165534-Planar-silicon-solar-cell}.
  [Accessed 21-01-2026].

\bibitem{shukla_aluminum_2025}
S.~Shukla and P.~Arora, \enquote{Aluminum as a competitive plasmonic material
  for the entire electromagnetic spectrum: {A} review,}
  {\protect\JournalTitle{Results in Optics}} \textbf{18}, 100760 (2025).

\bibitem{li_research_surfacePlasmon_2025}
T.~Li and J.~Zhu, \enquote{Research of {Enhancing} the {Light} {Absorption} and
  {Strengthen} {GaAs} {Thin} {Film} {Solar} {Cells} with {Metallic} {Ti}
  {Material} {Based} on {Grating} {Structure} and {Surface} {Plasmon}
  {Resonance} {Effect},} {\protect\JournalTitle{Plasmonics}}  (2025).

\bibitem{sultan2023TENCON}
R.~B. Sultan, A.~Al~Suny, S.~Tohfa, \emph{et~al.}, \enquote{Use of nano-grating
  structures embedded within the absorbing substrate to optimize the efficiency
  of cadmium telluride thin-film solar cells,} in \emph{TENCON 2023-2023 IEEE
  Region 10 Conference (TENCON),}  (IEEE, 2023), pp. 1234--1239.

\bibitem{optimization_utility}
\enquote{{O}ptimization utility --- optics.ansys.com,}
  \url{https://optics.ansys.com/hc/en-us/articles/360034922953-{O}ptimization-utility}.
  [Accessed 29-04-2026].

\bibitem{storn_differential_1997}
R.~Storn and K.~Price, \enquote{Differential {Evolution} – {A} {Simple} and
  {Efficient} {Heuristic} for global {Optimization} over {Continuous}
  {Spaces},} {\protect\JournalTitle{Journal of Global Optimization}}
  \textbf{11}, 341--359 (1997).

\bibitem{qiang_unified_2014}
J.~I. Qiang, \enquote{A {Unified} {Differential} {Evolution} {Algorithm} for
  {Global} {Optimization},} Tech. rep., Lawrence Berkeley National Laboratory
  (2014).

\bibitem{tsallis_generalized_1996}
C.~Tsallis and D.~A. Stariolo, \enquote{Generalized {Simulated} {Annealing},}
  {\protect\JournalTitle{Physica A: Statistical Mechanics and its
  Applications}} \textbf{233}, 395--406 (1996). ArXiv:cond-mat/9501047.

\bibitem{virtanen}
P.~Virtanen, R.~Gommers, T.~E. Oliphant, \emph{et~al.}, \enquote{{{SciPy} 1.0:
  Fundamental Algorithms for Scientific Computing in Python},}
  {\protect\JournalTitle{Nature Methods}} \textbf{17}, 261--272 (2020).

\bibitem{cb_honsberg_pveducation_nodate}
{C.B. Honsberg} and {S.G. Bowden}, \enquote{{PVEducation},} .

\bibitem{ma_trapassisted_2023}
X.~Ma, R.~A.~J. Janssen, and G.~H. Gelinck, \enquote{Trap‐{Assisted} {Charge}
  {Generation} and {Recombination} in {State}‐of‐the‐{Art} {Organic}
  {Photodetectors},} {\protect\JournalTitle{Advanced Materials Technologies}}
  \textbf{8}, 2300234 (2023).

\bibitem{Luo2016}
X.~Luo, R.~Hu, S.~Liu, and K.~Wang, \enquote{Heat and fluid flow in high-power
  led packaging and applications,} {\protect\JournalTitle{Progress in Energy
  and Combustion Science}} \textbf{56}, 1–32 (2016).

\bibitem{Hasan2021}
S.~N. Hasan, A.~J. Haque, T.~A. Khan, and M.~H. Chowdhury, \enquote{Use of
  aluminum-silica core-shell plasmonic nanoparticles to enhance the
  opto-electronic performance of thin-film solar cells,} in \emph{2021 6th
  International Conference on Development in Renewable Energy Technology
  (ICDRET),}  (IEEE, 2021), p. 1–6.

\bibitem{yu_effects_2017}
P.~Yu, Y.~Yao, J.~Wu, \emph{et~al.}, \enquote{Effects of {Plasmonic} {Metal}
  {Core} -{Dielectric} {Shell} {Nanoparticles} on the {Broadband} {Light}
  {Absorption} {Enhancement} in {Thin} {Film} {Solar} {Cells},}
  {\protect\JournalTitle{Scientific Reports}} \textbf{7}, 7696 (2017).

\bibitem{Teixeira2023}
F.~L. Teixeira, C.~Sarris, Y.~Zhang, \emph{et~al.}, \enquote{Finite-difference
  time-domain methods,} {\protect\JournalTitle{Nature Reviews Methods Primers}}
  \textbf{3} (2023).

\bibitem{solar_cell_methodology}
\enquote{{S}olar cell methodology --- optics.ansys.com,}
  \url{https://optics.ansys.com/hc/en-us/articles/360042165634-{S}olar-cell-methodology}.
  [Accessed 04-05-2026].

\bibitem{moree_comparison_2022}
G.~Mörée and M.~Leijon, \enquote{Comparison of {Poynting}’s vector and the
  power flow used in electrical engineering,} {\protect\JournalTitle{AIP
  Advances}} \textbf{12} (2022). Publisher: AIP Publishing.

\bibitem{su_based_2021}
J.~Su, H.~Yang, Y.~Xu, \emph{et~al.}, \enquote{Based on {Ultrathin}
  {PEDOT}:{PSS}/c-{Ge} {Solar} {Cells} {Design} and {Their} {Photoelectric}
  {Performance},} {\protect\JournalTitle{Coatings}} \textbf{11}, 748 (2021).
  Publisher: MDPI AG.

\bibitem{Suny2026}
A.~A. Suny, T.~Noor, M.~H. Hossain, \emph{et~al.}, \enquote{Broadband light
  absorption in cadmium telluride thin-film solar cells via composite light
  trapping techniques,} {\protect\JournalTitle{Nanoscale Advances}}  (2026).

\bibitem{wang_design_2017}
J.~Wang, Z.~Xu, F.~Bian, \emph{et~al.}, \enquote{Design and analysis of light
  trapping in thin-film gallium arsenide solar cells using an efficient hybrid
  nanostructure,} {\protect\JournalTitle{Journal of Nanophotonics}}
  \textbf{11}, 1 (2017). Publisher: SPIE-Intl Soc Optical Eng.

\bibitem{kennedy1995particle}
J.~Kennedy and R.~Eberhart, \enquote{Particle swarm optimization,} in
  \emph{Proceedings of ICNN'95-international conference on neural networks,}
  vol.~4 (ieee, 1995), pp. 1942--1948.

\bibitem{eberhart_new_optimizer}
R.~Eberhart and J.~Kennedy, \enquote{A new optimizer using particle swarm
  theory,} in \emph{{MHS}'95. {Proceedings} of the {Sixth} {International}
  {Symposium} on {Micro} {Machine} and {Human} {Science},}  (IEEE, Nagoya,
  Japan), pp. 39--43.

\bibitem{Spall2003-xl}
J.~C. Spall, \emph{Introduction to stochastic search and optimization}, Wiley
  Series in Discrete Mathematics and Optimization (John Wiley \& Sons,
  Nashville, TN, 2003).

\bibitem{scipy_differential_evolution_nodate}
\enquote{differential\_evolution — {SciPy} v1.16.0 {Manual},} .

\bibitem{Van_Laarhoven1987-vj}
P.~J.~M. Van~Laarhoven and E.~H.~L. Aarts, \emph{Simulated annealing: Theory
  and applications}, Mathematics and Its Applications (Kluwer Academic,
  Dordrecht, Netherlands, 1987), 1987th ed.

\bibitem{xin-she_yang_introduction_2019}
{Xin-She Yang}, \emph{Introduction to {Algorithms} for {Data} {Mining} and
  {Machine} {Learning}} (Elsevier, 2019).

\bibitem{afnan_uzzaman_sheikh_investigating_2024}
A.~F.~M. Afnan Uzzaman~Sheikh, A.~Karim, S.~A. Chowdhury, and M.~H. Chowdhury,
  \enquote{Investigating the {Effects} of {Edge} {Sharpness} of {Triangular}
  {Prism}-{Shaped} {Plasmonic} {Nanoparticles} on {Confining} {Light} to
  {Improve} the {Opto}-{Electronic} {Performance} of {Thin} {Film} {Solar}
  {Cells},} in \emph{2024 {International} {Conference} on {Sustainable}
  {Energy}: {Energy} {Transition} and {Net}-{Zero} {Climate} {Future}
  ({ICUE}),}  (IEEE, Pattaya City, Thailand, 2024), pp. 1--6.

\bibitem{chowdhury_optimization_2024}
S.~A. Chowdhury, A.~Karim, A.~F. M. A.~U. Sheikh, and M.~H. Chowdhury,
  \enquote{Optimization of {Clusters} of {Spherical} {Aluminum} {Nanoparticles}
  for {Enhancing} the {Efficiency} of {Thin} {Film} {Solar} {Cells},} in
  \emph{2024 6th {International} {Conference} on {Sustainable} {Technologies}
  for {Industry} 5.0 ({STI}),}  (IEEE, Narayanganj, Bangladesh, 2024), pp.
  1--6.

\bibitem{ok_continuous_2011}
J.~G. Ok, H.~J. Park, M.~K. Kwak, \emph{et~al.}, \enquote{Continuous
  {Patterning} of {Nanogratings} by {Nanochannel}‐{Guided} {Lithography} on
  {Liquid} {Resists},} {\protect\JournalTitle{Advanced Materials}} \textbf{23},
  4444--4448 (2011).

\bibitem{heidarzadeh_performance_2019}
H.~Heidarzadeh and A.~Tavousi, \enquote{Performance enhancement methods of an
  ultra-thin silicon solar cell using different shapes of back grating and
  angle of incidence light,} {\protect\JournalTitle{Materials Science and
  Engineering: B}} \textbf{240}, 1--6 (2019).

\bibitem{Arons1965}
A.~B. Arons and M.~B. Peppard, \enquote{Einstein’s proposal of the photon
  concept—a translation of the annalen der physik paper of 1905,}
  {\protect\JournalTitle{American Journal of Physics}} \textbf{33}, 367–374
  (1965).

\bibitem{zhu2017surface}
W.~Zhu, T.~Xu, H.~Wang, \emph{et~al.}, \enquote{Surface plasmon polariton laser
  based on a metallic trench fabry-perot resonator,}
  {\protect\JournalTitle{Science Advances}} \textbf{3}, e1700909 (2017).

\bibitem{prentice2000computer}
J.~Prentice, \enquote{Computer simulation of the effect of phosphorous doping
  of the i-layer in a thin-film a-si: H p--i--n solar cell,}
  {\protect\JournalTitle{Solar energy materials and solar cells}} \textbf{61},
  287--300 (2000).

\bibitem{kharb_modeling_2014}
R.~K. Kharb, S.~Shimi, S.~Chatterji, and M.~F. Ansari, \enquote{Modeling of
  solar {PV} module and maximum power point tracking using {ANFIS},}
  {\protect\JournalTitle{Renewable and Sustainable Energy Reviews}}
  \textbf{33}, 602--612 (2014).

\bibitem{li_aluminum_2020}
Z.~Li, C.~Li, J.~Yu, \emph{et~al.}, \enquote{Aluminum nanoparticle films with
  an enhanced hot-spot intensity for high-efficiency {SERS},}
  {\protect\JournalTitle{Optics Express}} \textbf{28}, 9174 (2020).

\bibitem{eze_optimum_2021}
M.~C. Eze, G.~Ugwuanyi, M.~Li, \emph{et~al.}, \enquote{Optimum silver contact
  sputtering parameters for efficient perovskite solar cell fabrication,}
  {\protect\JournalTitle{Solar Energy Materials and Solar Cells}} \textbf{230},
  111185 (2021).

\bibitem{arl_sio2_2020}
D.~Arl, V.~Rogé, N.~Adjeroud, \emph{et~al.}, \enquote{{SiO}$_{\textrm{2}}$
  thin film growth through a pure atomic layer deposition technique at room
  temperature,} {\protect\JournalTitle{RSC Advances}} \textbf{10}, 18073--18081
  (2020).

\bibitem{ping_pecvd_2011}
S.~Ping, L.~Jie, S.~Gao, \emph{et~al.}, \enquote{{PECVD} grown {SiO} 2 film
  process optimization,}  (San Francisco, California, 2011), p. 79431E.

\bibitem{seniutinas_nano-proximity_2016}
G.~Seniutinas, G.~Gervinskas, J.~Anguita, \emph{et~al.},
  \enquote{Nano-proximity direct ion beam writing,}
  {\protect\JournalTitle{Nanofabrication}} \textbf{2} (2016).

\bibitem{martin_fabrication_2015}
J.~Martin and J.~Plain, \enquote{Fabrication of aluminium nanostructures for
  plasmonics,} {\protect\JournalTitle{Journal of Physics D: Applied Physics}}
  \textbf{48}, 184002 (2015).

\bibitem{poitras_ion-beam_2003}
D.~Poitras, J.~A. Dobrowolski, T.~Cassidy, and S.~Moisa, \enquote{Ion-beam
  etching for the precise manufacture of optical coatings,}
  {\protect\JournalTitle{Applied Optics}} \textbf{42}, 4037 (2003).

\bibitem{liu_study_Si3n4_TFs_2013}
L.~Liu, W.-g. Liu, N.~Cao, and C.-l. Cai, \enquote{Study on {The} {Performance}
  of {PECVD} {Silicon} {Nitride} {Thin} {Films},}
  {\protect\JournalTitle{Defence Technology}} \textbf{9}, 121--126 (2013).

\bibitem{chen_growth_2020}
W.-C. Chen, S.~Chen, T.-Y. Yu, \emph{et~al.}, \enquote{Growth of {Si3N4} {Thin}
  {Films} on {Si}(111) {Surface} by {RF}-{N2} {Plasma} {Nitriding},}
  {\protect\JournalTitle{Coatings}} \textbf{11}, 2 (2020).

\bibitem{yoshinaga_low_2017}
S.~Yoshinaga, Y.~Ishikawa, and Y.~Uraoka, \enquote{Low surface reflectance by
  nanoimprinted texture with silicon-rich silicon nitride layer,}
  {\protect\JournalTitle{Journal of Physics D: Applied Physics}} \textbf{50},
  455108 (2017).

\bibitem{Mitrofanov2017}
M.~I. Mitrofanov, S.~N. Rodin, I.~V. Levitskii, \emph{et~al.}, \enquote{Ga
  focused ion beam etching of a si3n4/gan substrate for submicron selective
  epitaxy,} {\protect\JournalTitle{Journal of Physics: Conference Series}}
  \textbf{816}, 012009 (2017).

\end{thebibliography}


\begin{thebibliography}{1}
\newcommand{\enquote}[1]{``#1''}

\bibitem{ansysSelectingMesh}
\enquote{Selecting the best mesh refinement option in the fdtd simulation
  object --- optics.ansys.com,}
  \url{https://optics.ansys.com/hc/en-us/articles/360034382614-Selecting-the-best-mesh-refinement-option-in-the-FDTD-simulation-object}.
  [Accessed 20-01-2026].

\bibitem{Gedney2010}
S.~D. Gedney and B.~Zhao, \enquote{An auxiliary differential equation
  formulation for the complex-frequency shifted pml,}
  {\protect\JournalTitle{IEEE Transactions on Antennas and Propagation}}
  \textbf{58}, 838–847 (2010).

\bibitem{planar_si_sc}
\enquote{Planar silicon solar cell --- optics.ansys.com,}
  \url{https://optics.ansys.com/hc/en-us/articles/360042165534-Planar-silicon-solar-cell}.
  [Accessed 21-01-2026].

\end{thebibliography}

\end{document}


\maketitle

\section{Modeling and constraint details}
The nanostructure embedded within the silicon absorber layer is defined as a polygonal cross section which is extruded across the y-axis. As a result, a cross section of the nanostructure with arbitrary parameters can be used to show how the height of the rounded triangle can be equated to $0.5(3R + c)$. Fig. \ref{fig:height} is provided to illustrate how this can be deduced with simple trigonometry. It should be noted that when the corner radius $c$ is 0 the cross section is an equilateral triangle, thus each angle is 60$^\circ$. A derivation of the values shown in Fig. \ref{fig:height} is provided below. The lengths $R$, $a$ and the base of the inner equilateral triangle make a right-angled triangle, thus:

\begin{figure}[h]
\centering
\fbox{\includegraphics[width=.5\linewidth]{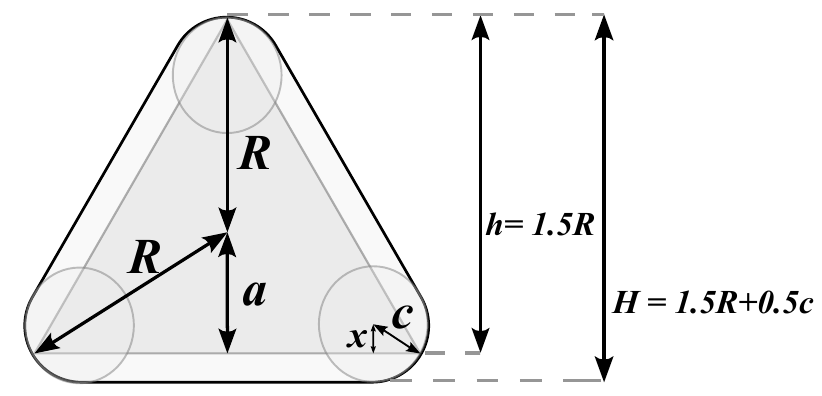}}
\caption{Geometry of cross-section of nanostructure depicted, showing how the full height can be found in terms of $R$ and $c$.}
\label{fig:height}
\end{figure}

\begin{equation*}
    \begin{split}
        \frac{a}{R} &= \sin(30^\circ) \\
        a &= 0.5R \\
    \end{split}
\label{}
\end{equation*}

\begin{equation}
    \therefore h = R+0.5R = 1.5R
\end{equation}

Building on this, the full height of the rounded triangle (denoted by $H$) can also be found with the simple derivation provided. 

\begin{equation*}
    \begin{split}
        \frac{x}{c} &= \sin(30^\circ) \\
        x &= 0.5c \\
    \end{split}
\label{}
\end{equation*}

\begin{equation}
    \therefore H = 1.5R + (c-0.5c) = 1.5R+0.5c = 0.5(3R+c) 
\end{equation}

\begin{figure}[h]
\centering
\fbox{\includegraphics[width=.6\linewidth]{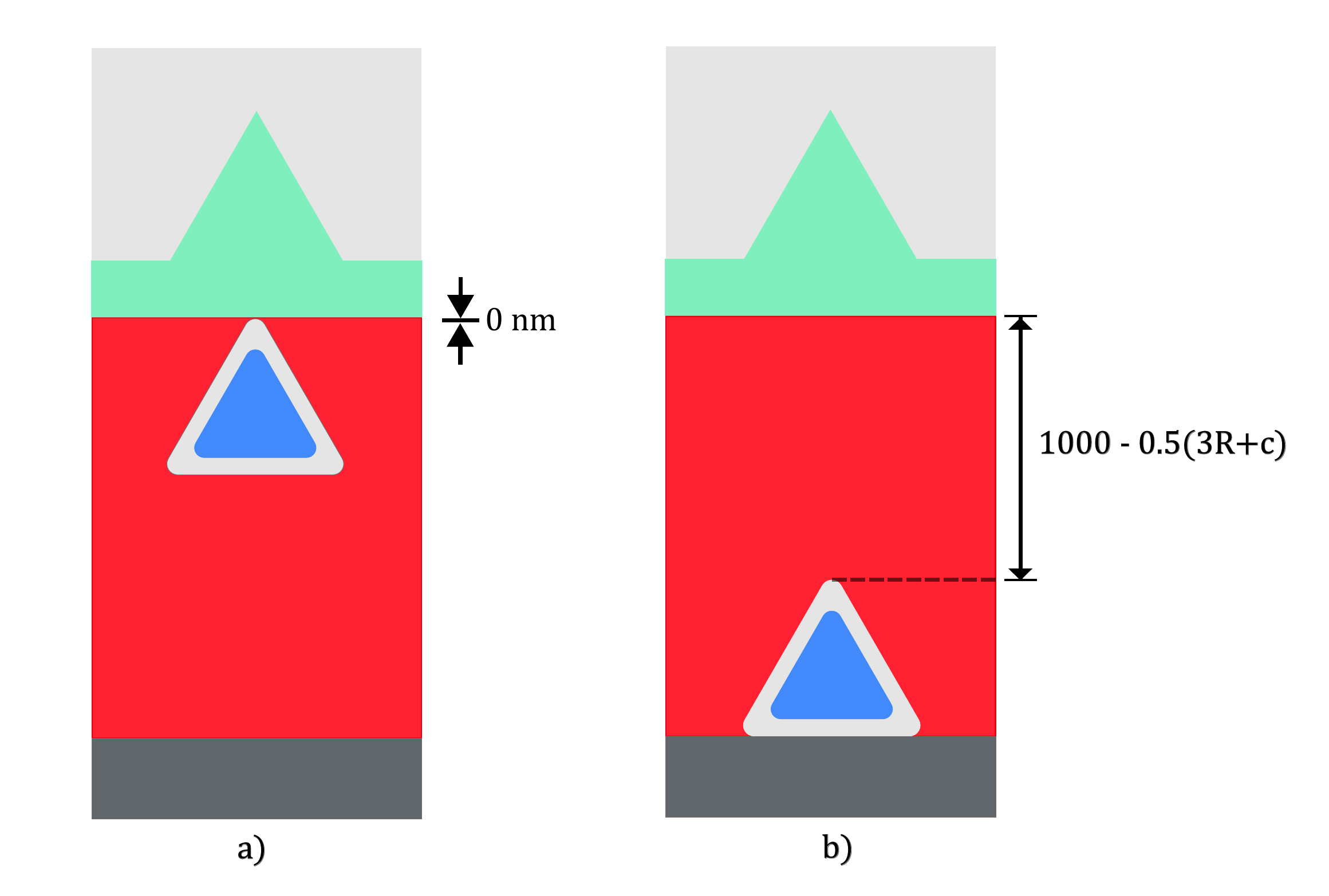}}
\caption{The minimum (a) and maximum (b) of the vertical displacement of the nanostructure are shown to illustrate the need for the conditional constraints, without which the geometry would overlap resulting in an unintended design.}
\label{fig:extzPos}
\end{figure}

As a result, the boundary values for the vertical position parameter, z, can be easily found as well. Fig. \ref{fig:extzPos} shows this to illustrate the edge cases for the vertical position, which need to be clearly defined to avoid unwanted design errors. Examples of unwanted design errors could include the nanostructure overlapping with the ARC layer or the Ag back reflector. As shown in Fig. \ref{fig:extzPos}(a), 0 nm indicates the highest position while Fig. \ref{fig:extzPos}(b) shows the lowest position, obtained by subtracting the height of the rounded triangular cross-section from the thickness of the silicon substrate $(1000-0.5(3R+c))$.

\section{Further parameter variation and trend analysis}

\begin{figure}[h]
\centering
\fbox{\includegraphics[width=1.0\linewidth]{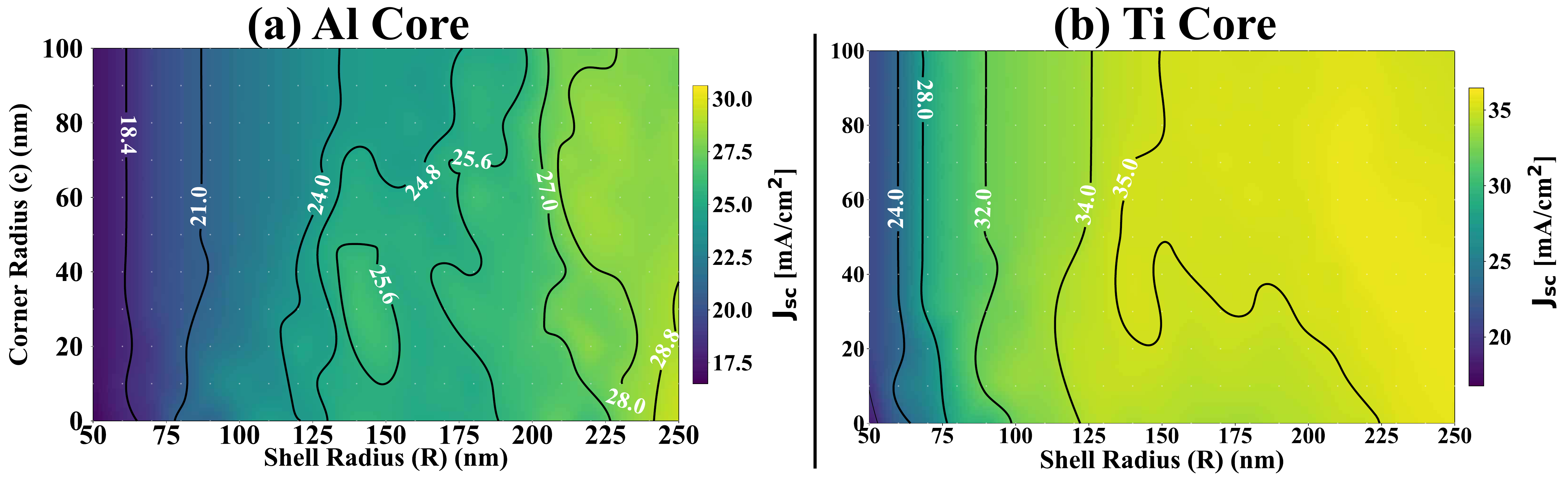}}
\caption{Contour plot where corner radius $(c)$ is varied against shell radius $(R)$  for embedded nanostructure and output J\textsubscript{SC} from FDTD simulations is recorded. (a) Al core variation with respect to $R$ and $c$. (b) Ti core variation with respect to $R$ and $c$.}
\label{fig:contour_cvR}
\end{figure}

\begin{figure}[h]
\centering
\fbox{\includegraphics[width=1.0\linewidth]{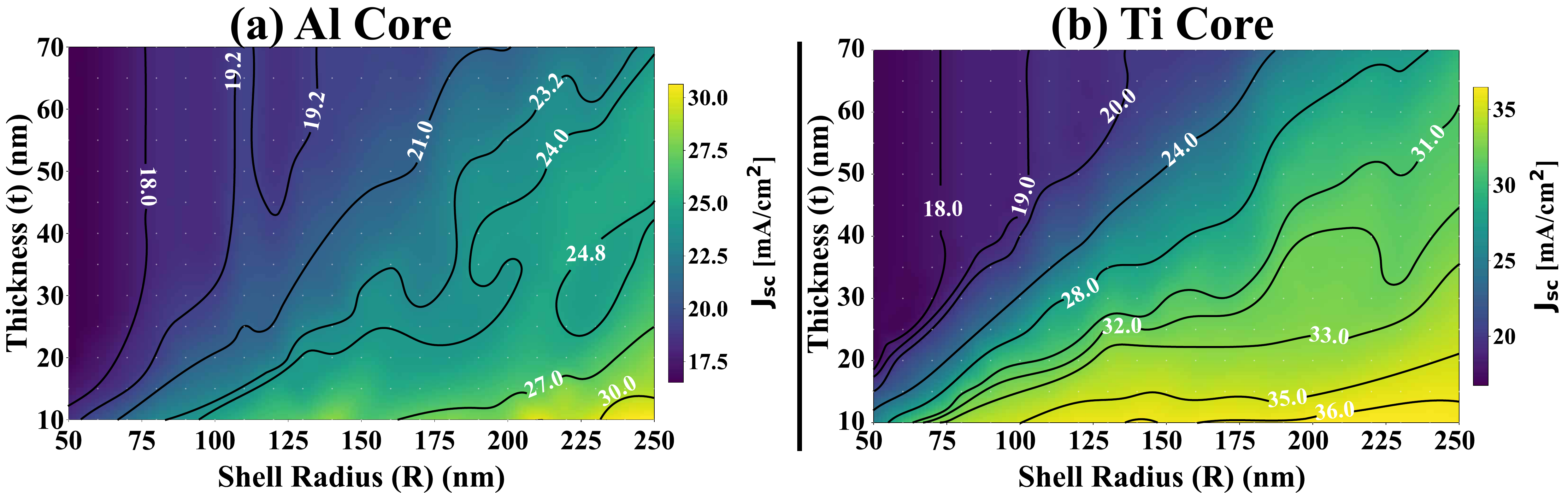}}
\caption{Contour plot where shell thickness $(t)$ is varied against shell radius $(R)$  for embedded nanostructure and output J\textsubscript{SC} from FDTD simulations are recorded. (a) Al core variation with respect to $R$ and $t$. (b) Ti core variation with respect to $R$ and $t$.}
\label{fig:contour_tvR}
\end{figure}

Fig. \ref{fig:contour_cvR} shows the variation of J\textsubscript{SC} when the radius $(R)$ and corner radius $(c)$ of the nanostructure are varied. For the nanostructure with Al core, shown in Fig. \ref{fig:contour_cvR}(a), lower values of $R$ generally lead to lower values of J\textsubscript{SC}, typically when $R$ is less than 100 nm. J\textsubscript{SC} increases with higher values of $R$ and the regions of highest J\textsubscript{SC} are found where $R$ is greater than 225 nm. In terms of corner radius ($c$), a clear change in J\textsubscript{SC} is not observed as it is changed. The contour plot implies that J\textsubscript{SC} mainly depends on $R$ in this case and is not much affected by $c$. The TFSC where the nanostructure core is Ti shows similar features (Fig. \ref{fig:contour_cvR}(b)). Again, low J\textsubscript{SC} regions are found where $R$ is less than 100 nm and J\textsubscript{SC} is barely affected by $c$. One difference that is more clearly seen in this figure than in the previous one is that the high J\textsubscript{SC} region starts where $c$ is higher. For lower values of $c$, the same J\textsubscript{SC} is generated when the value of $R$ is higher. This implies that while optimizing the nanostructure, similar output may be obtained for different input parameters. Fig. \ref{fig:contour_tvR} shows the variation in J\textsubscript{SC} when $R$ and $t$ are varied. Fig. \ref{fig:contour_tvR}(a) and (b) again show high regions at higher values of $R$. Generally, J\textsubscript{SC} increases as $t$ decreases for both Al and Ti cores. It can be inferred from Fig. \ref{fig:contour_tvR} that thinner dielectric shells may lead to better light trapping. A possible reason for this could be reduced parasitic absorption by the shell and easier transfer of energy by localized surface plasmon resonance (LSPR).

For Fig. \ref{fig:contour_cvR}, parameters $[t,p,z]$ are constant at $[15,100,500-R]$ while for \ref{fig:contour_tvR} the parameters $[c,p,z]$ are constant at $[20,50,500-R]$, all in nanometers. The $z$ parameter being fixed at $(500-R)$ is to keep the center of the nanostructure in place. From Fig. \ref{fig:contour_cvR} and \ref{fig:contour_tvR} along with Fig. 5, a qualitative conclusion can be made: that higher values of R, meaning larger nanostructures, are preferred if J\textsubscript{SC} is to be maximized. A general trend is that, for all cases,  Ti core outperformed Al as a core material when at maximizing J\textsubscript{SC}. The regions of high J\textsubscript{SC} formed for Ti core are much larger, meaning the J\textsubscript{SC} varies relatively less in response to changes in the nanostructure parameters. This capability is highly desirable when accounting for fabrication-induced variability, as it enables the device to maintain high performance levels even in the presence of structural imperfections. This is not the case with Al core as the regions of highest J\textsubscript{SC} values shown in Fig. 5(a), \ref{fig:contour_cvR}(a) and \ref{fig:contour_tvR}(a) are significantly smaller. Physically, this means that when TFSCs using each of the two different metal cores are fabricated, the TFSC with the Al core nanostructure will be more prone to producing lower performance than expected due to structural defects that can form during the fabrication process. Conversely, because the areas of highest output were larger for the Ti core, it can be said that even with fabrication defects as described within the scope of this study, the solar cell can be expected to maintain relatively high performance levels within a reasonable range of variation. However, because the other parameters were fixed for each of these variations, only a limited slice of the whole relation between input configuration parameters of the embedded nanostructures and the TFSC output is obtained.

\section{Further information regarding the optimal solution sensitivity analysis}

For the sensitivity analysis, the optimal parameters are varied to observe effects on the output. The range of variation here will be called the sub-range. The sub-range is a subset of the full optimization range. The variation sub-ranges for each optimal solution are provided below in Table \ref{tab:rangesData}. The corresponding optimal values for the sub-ranges shown in Table \ref{tab:rangesData} can be found in Tables 2 and 3 of the main paper. The sub-ranges for variation were determined based on the value of each parameter, following the set conditions described below. Generally, the bounds are the optimal value $\pm10\%$ of the optimization range. However, a condition is applied if the optimal parameter is too close to the upper or lower optimization bound. The variation sub-range cannot exceed the optimization bounds. If the difference between the optimal value and either bound of the optimization range is less than 10\% of the optimization range, then the upper or lower bound of the sub-range is clamped to the respective bound of the optimization range.

\begin{table}[h]
\caption{Variation ranges for each parameter for every solution found through optimization.}
\label{tab:rangesData}
\centering
\begin{tabular}  {|p{1.5cm}|p{1.5cm}|p{1.5cm}|p{1.5cm}|p{1.5cm}|p{1.5cm}|p{1.5cm}|}
\hline
Algorithm Solution & R variation (nm) & t variation (nm) & c variation (nm) & p variation (nm) & z variation (nm) \\ \hline
PSO-Al             & 230-250          & 10.0-16            & 40.7-78.2       & 0-57.5             & 0-59.5           \\ \hline
DE-Al              & 229-250          & 10.0-16.1            & 7.1-44.5         & 132-192          & 0-61.4           \\ \hline
DA-Al              & 227-250          & 10.0-17.0            & 0-31.5           & 138-198          & 0-62.3           \\ \hline
PSO-Ti             & 230-250          & 10.0-16.0            & 78.2-115.7       & 0-30             & 0-57.7           \\ \hline
DE-Ti              & 191-231          & 10.0-16.9            & 40.8-72.4        & 0-48.4             & 0-65.5           \\ \hline
DA-Ti              & 179-219          & 10.0-18.8            & 56.7-86.5        & 17.5-77.5        & 0-66.6  \\ \hline
\end{tabular}
\end{table}

The ranges calculated to test sensitivity were performed with the applied conditions. An arbitrary parameter $X_O$ can represent any optimized parameter from $[R,t,c,p,z]$. The terms $X_L$ and $X_U$ represent the lower and upper limits of the optimization range. Thus, the difference between them, denoted by T, can be used to find the length of this range with $T=X_U-X_L$. Generally, the bounds of the range for sampling can be found as $[X_O-0.1T,X_O+0.1T]$. However, if the optimal parameter is too close to the lower or upper bound of the optimization range, the sub-range must be clamped accordingly. For the first condition, if the following is true: 

\begin{equation}
|X_U-X_O|\leq0.1T
\label{eq:upCondition}
\end{equation}

Then the range is fixed with the intervals $[X_O-0.1T,X_U]$. Conversely, if the following condition in Eq. \ref{eq:lowCondition} is true:

\begin{equation}
|X_L-X_O|\leq0.1T
\label{eq:lowCondition}
\end{equation}

Then the range is fixed with the intervals $[X_L,X_O+0.1T]$.

\section{FDTD and CHARGE Simulation Details}

As shown in Fig. 1(a), the finite-difference time-domain (FDTD) region is 3D and cuboidal. The y-span is set to 10 nm to reduce simulation times; the z-span is 3300 nm and the x-span is a variable parameter because of periodicity $(p)$. The simulation time is 1000 fs, and the simulation takes place at 300 K by default. The background material is assumed to be either air or vacuum, with the real refractive index fixed at 1.0 with no extinction coefficient. The mesh type is set to auto non-uniform by default and the mesh refinement is also fixed to conformal variant 0 for a balance of accuracy and speed \cite{ansysSelectingMesh}. Regarding boundary conditions (BCs), the top and bottom must be perfectly matched layer (PML) and the remaining four are periodic (or symmetric and anti-symmetric). In the PML settings, "stretched coordinate PML" type, steep angle profile and 12 layers are used according to default settings \cite{Gedney2010}. A plane-wave source is used and the injection direction is the negative-z direction. Material properties for FDTD are implemented with $(n,k)$ refractive index values, where $n$ is the wavelength-dependent refractive index and $k$ is the extinction coefficient. The mesh order for each of the structures is set such that higher-priority structures are simulated over lower-priority ones.

\begin{table}[h]
\centering
\caption{Parameter values used for electrical and thermo-electrical simulations}
\label{tab:elecThermVariables}
\begin{tabular}{lp{4cm}p{4.5cm}}
\hline
\textbf{Parameters} & \textbf{Description} & \textbf{Nominal Values} \\
\hline
$\mu_n$ & Electron mobility & 1471 cm$^2$/V-s \\
$\mu_p$ & Hole mobility & 470.5 cm$^2$/V-s \\
$N_A$ & Uniform acceptor p-doping & $4.5 \times 10^{18} cm^{-3}$ \\
$N_D$ & Uniform donor n-doping & $4.5\times 10^{18} cm^{-3}$ \\
$T_n, T_p$ & Trap-assisted (Rsrh) carrier recombination lifetime & $3.3\times 10^{-6}$ , $4.0\times 10^{-6}$ s \\
$C_n, C_p$ & Auger carrier recombination & $2.8\times 10^{-31}$, $9.9\times 10^{-32}$ $cm^6/s$ \\
$C_{radiative}$ & Radiative recombination & $1.6\times 10^{-14}$ $cm^3/s$ \\
$T$ & Simulation temperature (no heat generation) & 300 K \\
$H_c$ & Thermal conductivity & 148 W/mK \\
$E_c$ & Electrical conductivity & $3.11\times 10^{-4}$ S/m \\
\hline
\end{tabular}
\label{tab:elecSimparams}
\end{table}

As for the CHARGE simulations, the specific material settings are specified according to Table \ref{tab:elecSimparams}. Factors such as charge carrier (electron and hole) mobility, doping concentration, recombination rates, temperature and conductivity are entered according to values applicable to silicon. This allows calculation of relevant electrical performance metrics.

\section{Electrical results considering temperature increase inside the solar cell due to heat generation}

\begin{table}[h]
\centering
\caption{CHARGE electrical results accounting for temperature increase inside the solar cell due to light and carrier generation}
\label{tab:thermElec}
\begin{tabular}{|p{1.7cm}|p{1.5cm}|p{1.5cm}|l|p{1.5cm}|p{1cm}|l|}
\hline
Temperature (K) & Algorithm-Core & J\textsubscript{SC} (mA/cm\textsuperscript{2}) & V\textsubscript{OC}  (V) & P\textsubscript{max} (mW/cm\textsuperscript{2}) & $\eta\  (\%)$   & FF     \\ \hline
317         & PSO Al         & 29.11 & 0.486 & 11.29 & 11.29 & 0.798 \\ \hline
315         & DE Al          & 29.44 & 0.488 & 11.47 & 11.47 & 0.799 \\ \hline
320         & DA Al          & 29.50 & 0.490 & 11.55 & 11.55 & 0.799 \\ \hline
308         & PSO Ti         & 35.42 & 0.491 & 13.95 & 13.95 & 0.801 \\ \hline
316         & DE Ti          & 34.65 & 0.491 & 13.61 & 13.61 & 0.800 \\ \hline
314         & DA Ti          & 34.60  & 0.490 & 13.57 & 13.57 & 0.800 \\ \hline
315         & No NS          & 13.67 & 0.464 & 4.90  & 4.90  & 0.773 \\ \hline
\end{tabular}
\end{table}

Table \ref{tab:thermElec} provides the data from electrical simulations, where the internal temperature of the TFSC is greater than 300 K due to heat generation.  The heat profile generated is due to the effect of light and the movement of charge carriers in the absorber layer \cite{planar_si_sc}. The temperature column represents the average steady-state temperature within the silicon of the TFSC while the cell is operating. Furthermore, Table \ref{tab:deltaElec} contains the difference in results between Table \ref{tab:thermElec} and Table 4 from the main manuscript to allow for a clear comparison. For clarity, the results in Table \ref{tab:deltaElec} are produced by subtracting the respective results in Table 4 from Table \ref{tab:thermElec}. As observed in Table \ref{tab:deltaElec}, there are no changes in V\textsubscript{OC} and FF. However, minor reductions in J\textsubscript{SC}, P\textsubscript{max}, and efficiency ($\eta$) can be observed overall, apart from two exceptions (PSO-Al and PSO-Ti).

\begin{table}[h]
\centering
\caption{Difference in electrical results from CHARGE simulations between results where heating effect is considered (>300 K) and when it is not considered (=300 K).}
\label{tab:deltaElec}
\begin{tabular}{|p{1.5cm}|p{1.3cm}|p{1.1cm}|l|p{1.5cm}|l|p{1.75cm}|}
\hline
Algorithm-Core & $\Delta$ J\textsubscript{SC} (mA/cm\textsuperscript{2}) & $\Delta$ V\textsubscript{OC} (V) & $\Delta$ FF & $\Delta$ P\textsubscript{max} (mW/cm\textsuperscript{2}) & $\Delta$ $\eta$ (\%) & $\Delta$ Temperature (K) \\ \hline
PSO-Al         & 0.22         & 0.000        & 0.000       & 0.09          & 0.09       & 17            \\ \hline
DE-Al          & -0.03         & 0.000        & 0.000       & -0.01          & -0.01       & 15            \\ \hline
DA-Al          & -0.09         & 0.000        & 0.000       & -0.05          & -0.05       & 20            \\ \hline
PSO-Ti         & 0.03         & 0.000        & 0.000       & 0.01          & 0.01       & 8             \\ \hline
DE-Ti          & -0.08         & 0.000        & 0.000       & -0.04          & -0.04       & 16            \\ \hline
DA-Ti          & -0.01         & 0.000        & 0.000       & 0.00          & 0.00       & 14            \\ \hline
No NS          & -0.02         & 0.000        & 0.000       & -0.01          & -0.01       & 15            \\ \hline
\end{tabular}
\end{table}

\bibliography{supp-journalBib}
